\documentclass[10pt,twocolumn,english,aps,prd,nofootinbib,preprintnumbers]{revtex4}
\usepackage[latin9]{inputenc}
\usepackage{graphicx}
\usepackage{amsmath}
\usepackage{amssymb}
\usepackage{amsfonts}
\usepackage{lscape}
\usepackage{hyperref}
\usepackage{url}
\usepackage{color}
\usepackage{bm}
\usepackage{tabularx}
\usepackage{mathtools}
\newcommand{\be}{\begin{equation}}
\newcommand{\ee}{\end{equation}}
\newcommand{\ba}{\begin{eqnarray}}
\newcommand{\ea}{\end{eqnarray}}
\newcommand{\nn}{\nonumber}

\renewcommand{\[}{\begin{equation}}
\renewcommand{\]}{\end{equation}}
\def\be{\begin{equation}}
\def\ee{\end{equation}}
\def\bea{\begin{eqnarray}}
\def\eea{\end{eqnarray}}
\def\eqi{\begin{equation}}
\def\eqf{\end{equation}}
\def\eqia{\begin{eqnarray}}
\def\eqfa{\end{eqnarray}}
\def\lcdm{$\Lambda$CDM }

\makeatother

\begin{document}

\preprint{IFT-UAM/CSIC-18-108}

\title{Unraveling the effective fluid approach for $f(R)$ models in the subhorizon approximation}

\author{Rub\'{e}n Arjona}
\email{ruben.arjona@estudiante.uam.es}

\author{Wilmar Cardona}
\email{wilmar.cardona@uam.es}

\author{Savvas Nesseris}
\email{savvas.nesseris@csic.es}

\affiliation{Instituto de F\'isica Te\'orica UAM-CSIC, Universidad Auton\'oma de Madrid,
Cantoblanco, 28049 Madrid, Spain}

\date{\today}

\begin{abstract}
We provide explicit formulas for the effective fluid approach of $f(R)$ theories, such as the Hu \& Sawicki and the designer models. Using the latter and simple modifications to the CLASS code, which we call EFCLASS, in conjunction with very accurate analytic approximations for the background evolution, we obtain competitive results in a much simpler and less error-prone approach. We also derive the initial conditions in matter domination and we find they differ from those already found in the literature for a constant $w$ model. A clear example is the designer model that behaves as $\Lambda$CDM in the background, but has nonetheless dark energy perturbations. We then use the aforementioned models to derive constraints from the latest cosmological data, including  supernovae, BAO, CMB, $H(z)$ and growth-rate data, and find they are statistically consistent to the $\Lambda$CDM model. Finally, we show that the viscosity parameter $c_{vis}^2$ in realistic models is not constant as commonly assumed, but rather evolves significantly over several orders of magnitude, something which could affect forecasts of upcoming surveys.
\end{abstract}
\maketitle

\section{Introduction}
\label{Section:Introduction}

A few decades ago it became clear that a model of the Universe including the cosmological constant $\Lambda$ could alleviate several problems in the Cold Dark Matter (CDM) scenario \cite{Kofman:1985fp}. Although the standard model of cosmology \lcdm is in very good agreement with recent astrophysical measurements \cite{Aghanim:2018eyx,Abbott:2017wau}, it is also well known that the huge discrepancy between both predicted and inferred values of $\Lambda$ represents one of the biggest conundrums for fundamental physics \cite{Weinberg:1988cp,Carroll:2000fy}.

In 1998 convincing evidence from observations of Supernovae type Ia (SnIa) showed that the Universe is undergoing a phase of accelerated expansion \cite{Riess:1998cb,Perlmutter:1998np}. Ever since, the standard cosmological model \lcdm has become the best phenomenological description for the Universe \cite{Hinshaw:2012aka,Aghanim:2018eyx,Abbott:2017wau}. The yet unsolved cosmological constant problem has driven an effort towards alternative explanations for the late-time accelerating phase of the Universe.

Different cosmological models have emerged and nowadays one finds two leading approaches which avoid the introduction of a cosmological constant. On the one hand, there exist Dark Energy (DE) models \cite{Copeland:2006wr} where yet unobserved scalar fields would dominate the energy content at late times, avoiding fine-tuning issues as well as accelerating the Universe \cite{Ratra:1987rm,ArmendarizPicon:2000dh}. On the other hand, there are Modified Gravity (MG) models that instead modify the current theory of gravity, namely, Einstein's Theory of General Relativity (GR) \cite{Clifton:2011jh}. These modifications of GR are however not easily achieved as several tests carried out up to extragalactic scales are in very good agreement with GR \cite{Collett:2018gpf,PhysRevLett.116.221101}.

Both DE and MG models provide plausible, alternative scenarios for explaining the late-time acceleration of the Universe. It is known that both kinds of models can fit background astrophysical observations, as well as the standard model $\Lambda$CDM. These models are therefore degenerated at the background level despite several efforts to disentangle them with model independent approaches \cite{Nesseris:2010ep,Nesseris:2012tt}. Although the recent discovery of gravitational waves by the LIGO Collaboration \cite{Abbott:2017oio} allows us to rule out some families of MG models \cite{Creminelli:2017sry,Sakstein:2017xjx,Ezquiaga:2017ekz,Baker:2017hug,Amendola:2017orw,Crisostomi:2017pjs,Frusciante:2018,Kase:2018aps,McManus:2016kxu,Lombriser:2015sxa} (e.g., from the so-called Horndeski theories\footnote{However, a recent work claims that the reduction of viable MG models is not as severe as previously announced \cite{Copeland:2018yuh}.} \cite{Horndeski:1974wa}), there remains a degeneracy between the two leading approaches.

Among the remaining MG models one finds an important class: $f(R)$ models \cite{Sotiriou:2008rp,DeFelice:2010aj,Nojiri:2017ncd,Nojiri:2010wj}. Even though this kind of model might be fully degenerated at the background level (e.g., the so-called designer $f(R)$ models which can exactly mimic the background dynamics of a dark energy model with equation of state $w(z)$ \cite{Multamaki:2005zs,delaCruzDombriz:2006fj,Pogosian:2007sw,Nesseris:2013fca}), the linear order perturbations could in principle be distinguishable from \lcdm \cite{Tsujikawa:2007gd}. This is relevant as in general the DE perturbations can have a strong effect in the determination of the growth-index $\gamma$ \cite{Nesseris:2015fqa}, even though with current growth data it is not possible to draw definite conclusions in favor of any $f(R)$ model \cite{Luna:2018tot,Perez-Romero:2017njc}.

The study of perturbations in MG models is thus of great importance and one can find different approaches in the literature (e.g., \cite{Hu:2007nk,Hu:2007pj,Tsujikawa:2007gd,Kunz:2006ca,Pogosian:2007sw,Koivisto:2006xf,Koivisto:2006ai,delaCruzDombriz:2008cp,delaCruzDombriz:2006fj,PhysRevD.76.104043,Starobinsky:2007hu,Bean:2006up,Song:2010rm,Pogosian:2010tj,Bean:2010zq,Caldwell:2007cw,Bertschinger:2008zb,PhysRevD.84.124018,Silvestri:2013ne,Clifton:2018cef,Ishak:2018his}). In Ref. \cite{Zhao:2008bn} the authors restricted themselves to background histories consistent with a flat \lcdm model and parameterized changes in both Poisson and anisotropy equations via two functions $\mu(a,k)$ and $\gamma(a,k)$; these two functions take into account possible deviations from GR in the relation between the Newtonian potentials as well as the relation between the potentials and matter perturbations. The parametric functions were implemented in a modified version of the code CAMB\footnote{\url{https://camb.info/} } \cite{Lewis:1999bs} dubbed MGCAMB.\footnote{\url{http://aliojjati.github.io/MGCAMB/home.html}} Since these parameterizations are only valid at late times, in Ref. \cite{Hojjati:2011ix} the authors modified MGCAMB to introduce new parameterizations which are valid at all times. A drawback in this approach to perturbations in MG models is that it fixes the background to \lcdm while it is known that viable $f(R)$ models might differ from \lcdm at the background level (e.g., Hu-Sawicki model \cite{Hu:2007nk}).

A different approach to study perturbations in MG models was carried out in Ref.~\cite{He:2012wq} where the author studied perturbations in $f(R)$ models which exactly mimic the \lcdm background by using the full set of covariant cosmological perturbation equations; the author modified the publicly available code CAMB, implemented this approach, and released a code called FRCAMB. \footnote{\url{http://darklight.fisica.unimi.it/cosmonews/frcamb/}} In Ref.~\cite{Xu:2015usa} the author extended FRCAMB to take into account $f(R)$ models with a background different from $\Lambda$CDM; the code has not been released.

An Effective Field Theory (EFT) approach \cite{Gubitosi:2012hu} to DE and MG models was pursued in Ref. \cite{Hu:2013twa} where authors had into account a fairly general theory with unbroken symmetries and implemented it in a code called EFTCAMB\footnote{\url{http://eftcamb.org/}} (i.e., a modified version of CAMB). Although this approach does not use any quasi-static approximation and evolves the full dynamics of perturbations on linear scales, the mapping of specific models into an EFT formalism might be cumbersome.

The Planck Collaboration used MGCAMB and EFTCAMB in Ref. \cite{Ade:2015rim} to study cosmological constraints in both DE and MG models. Although the results somehow depend on which data sets are regarded as well as on some assumptions (e.g., the equation of state $w(a)$, the sound speed $c_s^2(a,k)$, the anisotropic stress $\pi(a,k)$), the authors did not find conclusive evidence for extensions to the standard model of cosmology.\footnote{However, in Ref. \cite{Li:2018tfg} authors found evidence for deviations of GR ($\gtrsim 3\sigma$) using various astronomical observations, including data from Planck.}

In Ref. \cite{Battye:2015hza} authors proposed the so-called Equation of State (EOS) approach for perturbations. In this approach $f(R)$ models can be expressed as a dark energy fluid at background and linearized perturbation order \cite{Kunz:2006ca,Pogosian:2010tj}, see also \cite{Capozziello:2005mj,Capozziello:2006dj,Capozziello:2018ddp}. The authors used an elegant gauge-invariant formalism, without the sub-horizon approximation, where the modifications to GR are expressed as equation of state $w(a)$, entropy perturbation $\Gamma(a,k)$, and anisotropic stress $\Pi(a,k)$. The EOS approach was implemented in a modified version of the code CLASS\footnote{\url{http://class-code.net/}} \cite{Blas:2011rf} in Ref. \cite{Battye:2017ysh} where good agreement with previous studies and codes was found. In spite of addressing the problem of perturbations in $f(R)$ models in an elegant way, the EOS approach is not physically very intuitive: the interpretation of results and the perturbation variables in this formalism is not straightforward.

In this paper we will also express $f(R)$ models as a dark energy fluid, but differently to the EOS approach in \cite{Battye:2015hza}, as we will utilize the equation of state $w(a)$, the sound speed $c_s^2(a,k)$ and the anisotropic stress $\pi(a,k)$ as variables describing the fluid \cite{Kunz:2012aw}. This makes the comparison with popular DE models such as quintessence ($w(a)\geq -1$, $c_s^2=1$, $\pi(a,k)=0$) and K-essence ($w(a)$, $c_s^2(a)$, $\pi(a,k)=0$) relatively easy. This is of paramount importance in the case of the anisotropic stress because in $f(R)$ models generically one has $\pi(a,k) \neq 0$ whereas in standard single-field DE models $\pi(a,k) = 0$, so that any convincing evidence of anisotropic stress would rule out all standard single-field DE models \cite{PhysRevD.83.064042,Kunz:2012aw}. Likewise, non-detection of anisotropic stress would get several classes of MG models into difficulties.

Since current galaxy surveys do not reach scales comparable to the cosmological horizon, one frequently uses a quasi-static approximation for the perturbation equations. The quasi-static approximation roughly amounts to neglecting time derivatives in the linearized Einstein equations while only keeping spatial derivatives; in addition one only takes into account modes whose wavelength is shorter than the cosmological horizon. Some previous studies and implementations (i.e. FRCAMB, EFTCAMB, CLASS\_EOS\_FR) did not apply the sub-horizon approximation to the perturbation equations. Nevertheless, the quasi-static approximation has been investigated in the context of MG theories in Refs. \cite{delaCruzDombriz:2008cp,Sawicki:2015zya} and has been implemented in MGCAMB. On the one hand, in Ref. \cite{delaCruzDombriz:2008cp} authors argue that general $f(R)$ models do not satisfy the quasi-static approximation; however, the sub-horizon approximation can be safely used in $f(R)$ models describing the current phase of accelerating expansion and fulfilling solar system tests.

On the other hand, in Ref. \cite{Sawicki:2015zya} authors argue that the quasi-static approximation breaks down outside the DE sound-horizon $k\ll k_J$, where $k_J(z)\equiv \frac{H(z)}{(1+z) c_s}$ is the physical Jeans scale, rather than outside the cosmological horizon; the authors disregarded the anisotropic stress in their analysis and also consider a constant DE $c_s^2$, both assumptions being not realistic for viable MG models. In this paper we will work out solutions to the perturbations equations in $f(R)$ models under the sub-horizon approximation. We will derive analytical solutions for DE perturbations and test them numerically showing that the quasi-static approximation actually performs quite well for this kind of MG model.

By placing MG and DE models on the same framework one is, in principle, able to disentangle the two kinds of models through different predictions for the equation of state $w(a)$, the sound speed $c_s^2(a,k)$, and the anisotropic stress $\pi(a,k)$. Both DE sound speed and DE anisotropic stress are particularly important because they are closely related to the growth of structures and, therefore, might leave detectable traces in observables such as anisotropies in the Cosmic Microwave Background radiation (CMB) and Galaxy Counts (GC) \cite{Tsujikawa:2007gd,Cardona:2014iba}. Although DE and Dark Matter (DM) perturbations are invisible, they affect both the CMB and the GC via, for instance, the integrated Sachs-Wolfe (ISW) effect and the lensing potential \cite{Kunz:2006ca}. While the presence of DE anisotropic stress can enhance and stabilize the growth of matter perturbations \cite{PhysRevD.83.064042,Koivisto:2005mm,Mota:2007sz,PhysRevD.85.123529,Cardona:2014iba}, the DE sound speed might alter the level of clustering and the evolution of matter perturbations \cite{Hu:1998kj,dePutter:2010vy,Batista:2017lwf}. These properties are very important because one can use them to break background level degeneracies among different models \cite{Lewis:2002ah,Tegmark:2003ud}.

The most recent CMB data from the Planck satellite\footnote{\url{http://sci.esa.int/planck/}} as well as data from the Dark Energy Survey\footnote{\url{https://www.darkenergysurvey.org/}} are in good agreement with the standard cosmological model \lcdm \cite{Aghanim:2018eyx,Abbott:2017wau}, but this situation could potentially change by combining different probes and from upcoming galaxy surveys, stage IV CMB experiments, and gravitational wave observations (see, for instance, Refs. \cite{Amendola:2007rr,Amendola:2017ovw,Linder:2018jil,Peel:2018aly,Sapone:2013wda,Daniel:2010yt,Song:2010fg,PhysRevD.87.023501,Saltas:2014dha,Amendola:2014wma,Amendola:2013qna,PhysRevD.91.061501,Bonvin:2018ckp,Hagstotz:2018onp,Barbosa:2018iiq,Kase:2018aps,Linder:2018pth,Peel:2018aei,Zaldarriaga:1997ch,Bond:1997wr,Abbott:2017smn,Song:2010fg,Linder:2002et}). Despite the success of the \lcdm model when fitting current data sets, its Bayesian evidence\footnote{See, for instance, Refs. \cite{Marshall:2004zd,Trotta:2008qt} for a discussion about Bayesian evidence in cosmology.} is not extremely different from extended models \cite{Heavens:2017hkr,PhysRevD.98.063508}. Furthermore, there remain unexplained issues with other data sets such as direct Hubble constant measurements, weak lensing data, and cluster counts where dynamically DE models or MG models could play a part (see, for instance, Refs. \cite{Ade:2015xua,Zhao:2017cud,Heavens:2017hkr,Freedman:2017yms,Renk:2017rzu,Nunes:2018xbm,Lin:2018nxe,Cardona:2014iba,Benetti:2018zhv,PhysRevD.97.123504,Sakr:2018new}).

This paper is organized as follows. In Sec.~\ref{Section:Theoretical-framework} we discuss the standard equations for perturbations in a Friedmann-Lemaitre-Robertson-Walker (FLRW) metric. First, in Subsection \ref{subsection:fr-efa}, we explain how $f(R)$ models can be mapped into a DE fluid and give analytical solutions for DE perturbations in general $f(R)$ models under the sub-horizon approximation. Secondly, we present results for some viable $f(R)$ models in Subsection \ref{subsection:specific-models-results}. In Sec.~\ref{Section:numerical-solution} we show that our analytical solutions derived using the sub-horizon are in very good agreement with a full numerical evolution of the perturbation equations. Furthermore, we compare our implementation in the CLASS code with available codes such as MGCAMB, CLASS\_EOS\_FR, and FRCAMB. In Sec.~\ref{Section:viscosity} we clarify and discuss some points about viscosity in viable $f(R)$ models. Then, in Sec.~\ref{Section:mcmc-results} we present cosmological constraints for a few MG models within our methodology by using a Monte Carlo Markov Chain (MCMC) approach. We conclude in Sec.~\ref{Section:conclusions} and give details about our analytical computations and CLASS implementation in Appendices \ref{Section:useful-formulae} and \ref{Section:class-implementation}, respectively.   

\section{Theoretical framework}
\label{Section:Theoretical-framework}

Let us assume that the Universe can be described at the background level by a FLRW metric, then in order to study the perturbations of various cosmological models, we consider the perturbed FRW metric, which in the conformal Newtonian gauge can be written as:
\be
ds^2=a(\tau)^2\left[-(1+2\Psi(\vec{x},\tau))d\tau^2+(1-2\Phi(\vec{x},\tau))d\vec{x}^2\right],
\label{eq:FRWpert}
\ee
where $\tau$ is the conformal time defined via $d\tau=dt/a(t)$ and we will follow the notation of Ref.~\cite{Ma:1995ey}.\footnote{In more detail, our conventions are: (-+++) for the metric signature, the Riemann and Ricci tensors are given by $V_{b;cd}-V_{b;dc}=V_a R^a_{bcd}$ and $R_{ab}=R^s_{asb}$, while the Einstein equations are $G_{\mu\nu}=+\kappa T_{\mu\nu}$ for $\kappa=\frac{8\pi G_N}{c^4}$ and $G_N$ is the bare Newton's constant. In what follows we will set the speed of light $c=1$.}

At this point we can assume an ideal fluid with an energy momentum tensor
\be
T^\mu_{\nu}=P\delta^\mu_{\nu}+(\rho+P)U^\mu U_\nu,\label{eq:enten}
\ee
where $\rho$, $P$ are the fluid density and pressure, while $U^\mu=\frac{dx^\mu}{\sqrt{-ds^2}}$ is its velocity four-vector given to first order by $U^\mu=\frac{1}{a(\tau)}\left(1-\Psi,\vec{u}\right)$, which as can easily be seen satisfies $U^\mu U_\mu=-1$. Furthermore, $\vec{u}=\dot{\vec{x}}$, where $\dot{f}\equiv\frac{df}{d\tau}$, and the elements of the energy momentum tensor to first order of perturbations are given by:
\bea
T^0_0&=&-(\bar{\rho}+\delta \rho), \label{eq:effectTmn1}\\
T^0_i&=&(\bar{\rho}+\bar{P})u_i,\\
T^i_j&=& (\bar{P}+\delta P)\delta^i_j+\Sigma^i_j, \label{eq:effectTmn}
\eea
where $\bar{\rho},\bar{P}$ are defined on the background and are functions of time only, while the perturbations $\delta \rho, \delta P$ are functions of $(\vec{x},\tau)$ and $\Sigma^i_j\equiv T^i_j-\delta^i_j T^k_k/3$ is an anisotropic stress tensor.

Then, assuming GR we find that the perturbed Einstein equations in the conformal Newtonian gauge are given by \cite{Ma:1995ey}:
\be
k^2\Phi+3\frac{\dot{a}}{a}\left(\dot{\Phi}+\frac{\dot{a}}{a}\Psi\right) = 4 \pi G_N a^2 \delta T^0_0, \label{eq:phiprimeeq}
\ee
\be
k^2\left(\dot{\Phi}+\frac{\dot{a}}{a}\Psi\right) = 4 \pi G_N a^2 (\bar{\rho}+\bar{P})\theta,\label{eq:phiprimeeq1}
\ee
\be
\ddot{\Phi}+\frac{\dot{a}}{a}(\dot{\Psi}+2\dot{\Phi})+\left(2\frac{\ddot{a}}{a}-
 \frac{\dot{a}^2}{a^2}\right)\Psi+\frac{k^2}{3}(\Phi-\Psi)
=\frac{4\pi}{3}G_N a^2\delta T^i_i,
\ee
\be
k^2(\Phi-\Psi) = 12\pi G_N a^2 (\bar{\rho}+\bar{P})\sigma \label{eq:anisoeq},
\ee
where we have defined the velocity $\theta\equiv ik^ju_j$, the anisotropic stress  $(\bar{\rho}+\bar{P})\sigma\equiv-(\hat{k}_i\hat{k}_j-\frac13 \delta_{ij})\Sigma^{ij}$. We also need the evolution equations for the perturbations, given by the energy-momentum conservation $T^{\mu\nu}_{;\nu}=0$ as:
\be
\dot{\delta} = -(1+w)(\theta-3\dot{\Phi})-3\frac{\dot{a}}{a}\left(c_s^2-w\right)\delta,
\label{eq:cons1}
\ee
\be
\dot{\theta} = -\frac{\dot{a}}{a}(1-3w)\theta-\frac{\dot{w}}{1+w}\theta+\frac{c_s^2}{1+w}k^2\delta-k^2\sigma+k^2\Psi,
\label{eq:cons2}
\ee
where we define the equation of state parameter $w\equiv\frac{\bar{P}}{\bar{\rho}}$ and the rest-frame sound speed of the fluid $c_s^2\equiv\frac{\delta P}{\delta \rho}$. Following Ref.~\cite{Cardona:2014iba}, we eliminate $\theta$ from Eqs.~(\ref{eq:cons1}) and (\ref{eq:cons2}), resulting in a second order equation for $\delta$:
\bea
\ddot{\delta}&+&(\cdots) \dot{\delta}+(\cdots) \delta = \nn \\
&-& k^2\left((1+w)\Psi+c_s^2\delta-(1+w)\sigma\right)+\cdots \nn \\
&=& -k^2\left((1+w)\Psi+c_s^2\delta-\frac23\pi\right)+\cdots,
\eea
where the $(\cdots)$ indicates the presence of complicated expressions and we have defined the anisotropic stress parameter of the fluid as $\pi\equiv\frac32(1+w)\sigma$. As also discussed in Ref.~\cite{Cardona:2014iba} the $k^2$ term will act as a source, driving the perturbations. However, since the potential scales as $\Psi\sim1/k^2$ in relevant scales, the only terms that matter are the sound speed and the anisotropic stress. Therefore, we can define an effective sound speed as
\be
c_{s,eff}^2 =c_s^2-\frac23\pi/\delta\label{eq:cs2eff}
\ee
that characterizes the propagation of perturbations as well as the clustering properties on sub-horizon scales. We should also note that in principle the sound speed $c_s^2$ can be both time and scale dependent, i.e., $c_s^2=c_s^2(\tau,k)$. For example, as noted in Ref.~\cite{Amendola:2015ksp}, the sound speed for a scalar field $\phi$ in the conformal Newtonian gauge for small scales is $c_{s,\phi}^2\simeq\frac{k^2}{4 a^2 m_\phi^2}$, where $m_\phi$ is the mass of the scalar field. On the other hand, $c_s^2$ is equal to one only in the scalar field's rest-frame (see Chapter 11.2 of Ref. \cite{Amendola:2015ksp} for a quick derivation). Of course, one has the same situation in $f(R)$ theories because in practice they only contain a scalar degree of freedom\footnote{$f(R)$ theories can be viewed as a non-minimally coupled scalar field in the Einstein frame. See, for instance, Ref. \cite{Mukhanov:1990me}.} \cite{Sawicki:2015zya}. Therefore, we expect the sound speed to be scale dependent in modified gravity models, when we are not in the rest frame of the equivalent DE fluid.

Finally, in what follows we will use the scalar velocity perturbation $V\equiv i k_jT^j_0/\rho=(1+w)\theta$ instead of the velocity $\theta$. The former has the advantage that it can remain finite when the equation of state $w$ of the fluid crosses $-1$ (see also Ref.~\cite{Sapone:2009mb}). With this new variable the evolution equations, Eqs. \eqref{eq:cons1}-\eqref{eq:cons2}, become

\bea
\delta' &=& 3(1+w) \Phi'-\frac{V}{a^2 H}-\frac{3}{a}\left(\frac{\delta P}{\bar{\rho}}-w\delta\right),
\label{Eq:evolution-delta}
\eea
\bea
V' &=& -(1-3w)\frac{V}{a}+\frac{k^2}{a^2 H}\frac{\delta P}{\bar{\rho}} +(1+w)\frac{k^2}{a^2 H} \Psi \nn \\
 &-&\frac23 \frac{k^2}{a^2 H} \pi,
\label{Eq:evolution-V}
\eea
where the prime $'$ is a derivative with respect to the scale factor $a$ and $H(t)=\frac{da/dt}{a}$ is the Hubble parameter.

\subsection{The f(R) models and the effective fluid approach}
\label{subsection:fr-efa}

In this set up we can study a plethora of MG models either directly as in Ref.~\cite{Tsujikawa:2007gd} or as an effective DE fluid \cite{Battye:2015hza}. For example, in the case of the $f(R)$ models, the modified Einstein-Hilbert action reads:
\be
S=\int d^{4}x\sqrt{-g}\left[  \frac{1}{2\kappa}f\left(  R\right)
+\mathcal{L}_{m}\right],  \label{eq:action1}%
\ee
where $\mathcal{L}_{m}$ is the Lagrangian of matter and
$\kappa=8\pi G_N$ is a constant with $G_N$ being the bare Newton's constant. Varying the action with respect to the metric, following the metric variational approach, we arrive at the following field equations  \cite{Tsujikawa:2007gd}:
\be
F G_{\mu\nu}-\frac12(f(R)-R~F) g_{\mu\nu}+\left(g_{\mu\nu}\Box-\nabla_\mu\nabla_\nu\right)F =\kappa\,T_{\mu\nu}^{(m)},
\label{eq:EE}
\ee
where $F=f'(R)$, $G_{\mu\nu}$ is the Einstein tensor and $T_{\mu\nu}^{(m)}$ is the energy-momentum tensor for the matter fields. By adding and subtracting the Einstein tensor on the left hand side of Eq.~(\ref{eq:EE}) and moving everything to the right hand side we can rewrite the equations of motion as the usual Einstein equations plus an effective DE fluid, along with the usual matter fields \cite{Pogosian:2010tj}:
\bea
G_{\mu\nu}&=&\kappa\left(T_{\mu\nu}^{(m)}+T_{\mu\nu}^{(DE)}\right),
\label{eq:effEqs}
\eea
where
\bea
\kappa T_{\mu\nu}^{(DE)}&=&(1-F)G_{\mu\nu}+\frac12(f(R)-R~F) g_{\mu\nu} \nn \\
 &-&\left(g_{\mu\nu}\Box-\nabla_\mu\nabla_\nu\right)F.
\label{eq:effTmn}
\eea
Due to the diffeomorphism invariance of the theory, it is very easy to show that the effective energy momentum tensor given by Eq.~\eqref{eq:effTmn}, indeed satisfies the usual conservation equation:
\be
\nabla^\mu T_{\mu\nu}^{(DE)}=0.
\ee
Clearly, the background equations are the same as in GR \cite{Ma:1995ey}:
\bea
\mathcal{H}^2&=&\frac{\kappa}{3}a^2 \left(\bar{\rho}_{m}+\bar{\rho}_{DE}\right), \\
\dot{\mathcal{H}}&=&-\frac{\kappa}{6}a^2 \left(\left(\bar{\rho}_{m}+3\bar{P}_{m}\right)+\left(\bar{\rho}_{DE}+3\bar{P}_{DE}\right)\right).
\eea
While we assume that matter is pressureless ($\bar{P}_{m}=0$), the effective DE density and pressure are given by:
\bea
\kappa \bar{P}_{DE}&=&\frac{f}2-\mathcal{H}^2/a^2-2F\mathcal{H}^2/a^2+\mathcal{H}\dot{F}/a^2 \nn \\
&-&2\dot{\mathcal{H}}/a^2-F\dot{\mathcal{H}}/a^2+\ddot{F}/a^2,\label{eq:effpr}\\
\kappa \bar{\rho}_{DE}&=&-\frac{f}2+3\mathcal{H}^2/a^2-3\mathcal{H}\dot{F}/a^2+3F\dot{\mathcal{H}}/a^2,\label{eq:effden}
\eea
where $\mathcal{H}=\frac{\dot{a}}{a}$ is the conformal Hubble parameter.\footnote{In what follows we denote the usual Hubble parameter as $H(t)=\frac{da/dt}{a}$ and the conformal one as $\mathcal{H}(\tau)=\frac{da/d\tau}{a}$. The two are related via $\mathcal{H}(\tau)=a H(t)$.}

Using Eqs.~(\ref{eq:effpr}) and (\ref{eq:effden}) we see that the DE equation of state for the $f(R)$ models in the effective fluid description is given by:
\be
w_{DE}=\frac{-a^2 f+2\left((1+2F)\mathcal{H}^2-\mathcal{H}\dot{F}+(2+F)\dot{\mathcal{H}}-\ddot{F}\right)}{a^2 f-6(\mathcal{H}^2-\mathcal{H}\dot{F}+F\dot{\mathcal{H}})}\label{eq:wde},
\ee
which is in agreement with the expression found in Ref.~\cite{Tsujikawa:2007gd}.

Thus it becomes clear that by working in the effective fluid approach, we can assign a density, pressure, velocity and anisotropic stress to the effective energy momentum tensor as in the general case of Eqs. \eqref{eq:effectTmn1}-\eqref{eq:effectTmn}. Then, we can find the effective quantities for the $f(R)$ model using the tensor of Eq.~(\ref{eq:effTmn}). As a result, the effective pressure, density and velocity perturbations are given by:
\bea
\frac{\delta P_{DE}}{\bar{\rho}_{DE}}&=&(...)\delta R+(...)\dot{\delta R}+(...)\ddot{\delta R}+(...)\Psi \nn \\
&+&(...)\dot{\Psi}+(...)\Phi+(...)\dot{\Phi},\label{eq:effdenp0}\\
\delta_{DE}&=&(...)\delta R+(...)\dot{\delta R}+(...)\Psi+(...)\Phi \nn \\
&+&(...)\dot{\Phi},\label{eq:effprp0} \\
V_{DE}&\equiv&(1+w_{DE})\theta_{DE} \nn \\
 &=&(...)\delta R+(...)\dot{\delta R}+(...)\Psi+(...)\Phi \nn \\
&+&(...)\dot{\Phi}.\label{eq:efftheta0}
\eea
Moreover, in these models it is easy to see from the field equations that the difference of the potentials $\Phi$ and $\Psi$ is given by
\be
\Phi-\Psi=\frac{F_{,R}}{F} \delta R,\label{eq:defaniso}
\ee
which implies that the anisotropic stress can be written as \cite{Ma:1995ey}
\bea
\bar{\rho}_{DE} \pi_{DE}&=&-\frac32 (\hat{k}_i\hat{k}_j-\frac13 \delta_{ij})\Sigma^{ij} \nn \\
&=&\frac{1}{\kappa}\frac{k^2}{a^2}\left(F_{,R} \delta R+(1-F)(\Phi-\Psi)\right).
\label{Eq:DE-rho-anisotropic-stress}
\eea
In Appendix \ref{Section:useful-formulae} we give some other useful expressions related to the effective fluid variables.

\subsubsection{Sub-horizon approximation}

Expressions in Eqs. \eqref{eq:effdenp0}-\eqref{Eq:DE-rho-anisotropic-stress} for DE perturbations might be cumbersome. Therefore, it is very convenient to work in the sub-horizon approximation, i.e., with modes deep in the Hubble radius $(k^2\gg a^2 H^2)$, where we find that terms with time-derivatives are negligible compared to the ones scaling as $k^2$. For example, the perturbation in the Ricci scalar is
\bea
\delta R&=&-\frac{12  (\mathcal{H}^2+\dot{\mathcal{H}})}{a^2}\Psi-\frac{4 k^2}{a^2}\Phi+\frac{2 k^2 }{a^2}\Psi \nn \\
&-&\frac{18 \mathcal{H} }{a^2}\dot{\Phi}-\frac{6 \mathcal{H} }{a^2}\dot{\Psi}-\frac{6 \ddot{\Phi}}{a^2},\label{eq:ricciexact}\\
&\simeq & -\frac{4 k^2}{a^2}\Phi+\frac{2 k^2}{a^2}\Psi,\label{eq:ricciapp}
\eea
where the last line follows from the sub-horizon approximation. Then, using the equations of motion we find that the potentials can be written as:
\bea
\Psi &=& -4\pi G_N \frac{a^2}{k^2} \frac{G_{eff}}{G_N} \bar{\rho}_m \delta_m, \label{eq:pot1} \\
\Phi &=& -4\pi G_N \frac{a^2}{k^2} Q_{eff} \bar{\rho}_m \delta_m,\label{eq:pot2}
\eea
where the effective Newton's constant $G_{eff}$ and $Q_{eff}$ are given by \cite{Tsujikawa:2007gd}:
\bea
G_{eff}/G_N&=& \frac{1}{F} \frac{1+4\frac{k^2}{a^2}\frac{F_{,R}}{F}}{1+3\frac{k^2}{a^2}\frac{F_{,R}}{F}}, \label{eq:pot1} \\
Q_{eff}&=& \frac{1}{F}  \frac{1+2\frac{k^2}{a^2}\frac{F_{,R}}{F}}{1+3\frac{k^2}{a^2}\frac{F_{,R}}{F}},\label{eq:pot2}
\eea
where $F=\frac{d f(R)}{dR}$, $F_{,R}=\frac{d^2 f(R)}{dR^2}$. Note however, that in the effective fluid approach we have to introduce the DE density $\rho_{DE}$, which then means that from the Poisson equation for $\Phi$ we have:
\bea
-\frac{k^2}{a^2}\Phi&=&4 \pi G_N\left( \bar{\rho}_m \delta_m+\bar{\rho}_{DE}\delta_{DE}\right)\nn \\
&=& 4 \pi G_N Q_{eff} \bar{\rho}_m \delta_m,
\eea
or that
\be
\bar{\rho}_m \delta_m=\frac{1}{Q_{eff}-1}\bar{\rho}_{DE}\delta_{DE},\label{eq:rmdm}\ee
which can be used to find the evolution of the DE density perturbation in this regime.

The previous expressions are also useful as in the sub-horizon approximation one can derive a second order differential equation for the matter density contrast in terms of $G_{eff}$ \cite{Tsujikawa:2007gd}:
\be
\delta_m''(a)+\left(\frac{3}{a}+\frac{H'(a)}{H(a)}\right)\delta_m'(a)-\frac32 \frac{\Omega_{m0} G_{eff}/G_N}{a^5 H(a)^2/H_0^2}\delta_m(a)=0, \label{eq:Geffode}
\ee
where in this case primes $'$ denote derivatives with respect to the scale factor $a$.

Finally, we can also define the anisotropic parameters $\eta\equiv \frac{\Psi-\Phi}{\Phi}$ and $\gamma\equiv\frac{\Phi}{\Psi}$ for which we then have,
\bea
\eta&=& \frac{2\frac{k^2}{a^2}\frac{F_{,R}}{F}}{1+2\frac{k^2}{a^2}\frac{F_{,R}}{F}}, \label{eq:an1} \\
\gamma &=&\frac{1+2\frac{k^2}{a^2}\frac{F_{,R}}{F}}{1+4\frac{k^2}{a^2}\frac{F_{,R}}{F}}. \label{eq:an2}
\eea

We can now apply the sub-horizon approximation and derive relatively simple expressions for all the effective DE perturbations in Eqs.~\eqref{eq:effdenp0}-\eqref{Eq:DE-rho-anisotropic-stress}. In practice, we have found that the results depend on the way the approximation is applied and this is one of the main results of our paper.

Since $\delta R$ in Eq. \eqref{eq:ricciexact} has up to second order derivatives of $\Phi$ and Eq.~(\ref{eq:effdenp0}) contains up to second order derivatives of $\delta R$, this means that the pressure perturbation has up to fourth order derivatives of the metric perturbation $\Phi$. Eliminating all of the higher order perturbations via the sub-horizon approximation can cause significant deviations and instabilities in the system of effective fluid equations. We found that a better approach is to use Eq.~\eqref{eq:effdenp0} and repeatedly apply Eq.~(\ref{eq:defaniso}), thus reducing the number of higher order derivative terms and increasing the accuracy of the solutions.

Following this prescription and using the Poisson equations for the potentials, we find that the effective density, pressure and velocity perturbations are given by:
\bea
\frac{\delta P_{DE}}{\bar{\rho}_{DE}}&\simeq&\frac{1}{3F}\frac{2\frac{k^2}{a^2}\frac{F_{,R}}{F}+3(1+5\frac{k^2}{a^2}\frac{F_{,R}}{F})\ddot{F}k^{-2}}{1+3\frac{k^2}{a^2}\frac{F_{,R}}{F}}\frac{\bar{\rho}_m}{\bar{\rho}_{DE}} \delta_m,\label{eq:effpres} \nn \\
& & \\
\delta_{DE}&\simeq&\frac{1}{F}\frac{1-F+\frac{k^2}{a^2}(2-3F)\frac{F_{,R}}{F}}{1+3\frac{k^2}{a^2}\frac{F_{,R}}{F}}\frac{\bar{\rho}_m}{\bar{\rho}_{DE}} \delta_m,\label{eq:effder} \\
V_{DE}&\equiv& (1+w_{DE})\theta_{DE}\nn \\ &\simeq&\frac{\dot{F}}{2F}\frac{1+6\frac{k^2}{a^2}\frac{F_{,R}}{F}}{1+3\frac{k^2}{a^2}\frac{F_{,R}}{F}}\frac{\bar{\rho}_m}{\bar{\rho}_{DE}} \delta_m.\label{eq:efftheta}
\eea
Finally, the DE anisotropic stress parameter $\pi_{DE}$ is given by
\bea
\pi_{DE}&=& \frac{\frac{k^2}{a^2} (\Phi-\Psi)}{\kappa~ \bar{\rho}_{DE}}\nn\\
&\simeq& \frac{1}{F}\frac{\frac{k^2}{a^2}\frac{F_{,R}}{F}}{1+3\frac{k^2}{a^2}\frac{F_{,R}}{F}}\frac{\bar{\rho}_m}{\bar{\rho}_{DE}} \delta_m \nn\\
&\simeq&\frac{\frac{k^2}{a^2}\frac{F_{,R}}{F}}{1-F+\frac{k^2}{a^2}(2-3F)\frac{F_{,R}}{F}}\delta_{DE}.\label{eq:effpi}
\eea
Note that the DE anisotropic stress in Eq. \eqref{eq:effpi} can also be written as
\bea
\pi_{DE}(a)&=& \frac{\frac{k^2}{a^2}f_1(a)}{1+\frac{k^2}{a^2}f_2(a)}\delta_{DE}(a),
\eea
where $f_1(a)=\frac{F_{,R}}{F (1-F)}$ and $f_2(a)=\frac{(2-3F)F_{,R}}{F (1-F)}$, which is reminiscent of Model 2 in Ref.~\cite{Cardona:2014iba}, but with different functions in the numerator and the denominator. This is interesting as it seems that many popular ansatze for the DE anisotropic stress do not capture exactly all of the features of the $f(R)$ models.

On the other hand, using Eqs.~(\ref{eq:effpres}) and (\ref{eq:effder}), we see that the DE sound speed is given by
\be
c_{s,DE}^2\simeq\frac13 \frac{2\frac{k^2}{a^2}\frac{F_{,R}}{F}+3(1+5\frac{k^2}{a^2}\frac{F_{,R}}{F})\ddot{F}k^{-2}}{1-F+\frac{k^2}{a^2}(2-3F)\frac{F_{,R}}{F}},
\label{eq:cs2de}
\ee
which implies that the DE effective sound speed is
\bea
c_{s,eff}^2 &\equiv& c_{s,DE}^2-\frac23\pi_{DE}/\delta_{DE}\nn \\
&\simeq& \frac{(1+5\frac{k^2}{a^2}\frac{F_{,R}}{F})\ddot{F}k^{-2}}{1-F+\frac{k^2}{a^2}(2-3F)\frac{F_{,R}}{F}}.\label{eq:cs21}
\eea
As we will see later on, the effective sound-speed at late times tends to go to zero due to the fact that the $\ddot{F}$ term not only is suppressed by $k^2$, which in the sub-horizon approximation is much larger than the Hubble parameter or related quantities, but also because for viable models $F$ in general is a slowly varying function. This implies that for these models there is no effective sound speed driving the DE perturbations, thus we expect that on large $k$ and at late times the perturbations should become flat, in agreement with Ref.~\cite{Cardona:2014iba}.

It is clear that for the \lcdm model, i.e., $f(R)=R-2\Lambda$, we have $F=1$ and $F_{,R}=0$ which implies that $w_{DE}=-1$ and $(\delta P_{DE},\delta \rho_{DE},\pi_{DE})=(0,0,0)$ as expected. When the equation of state $w_{DE}$ for an $f(R)$ model, e.g., the Hu \& Sawicki (HS, hereafter) model, crosses the so-called phantom divide line ($w_{DE}(a)=-1$), problems could arise due to the presence of the $1+w$ term in the denominator in Eq.~(\ref{eq:cons2}) \cite{Nesseris:2006er}. However, we see that in our case the perturbations remain finite despite the presence of the $1+w$ term in the denominator in Eq.~(\ref{eq:cons2}) as we can absorb the $1+w$ term by introducing $V_{DE}=(1+w_{DE})\theta_{DE}$ as mentioned earlier. Furthermore, the combination $(1+w_{DE})\theta_{DE}$ always remains finite for viable $f(R)$ models as can be seen in Eq.~(\ref{eq:efftheta}). The simple analytical expressions given by Eqs. \eqref{eq:effpres}-\eqref{eq:efftheta} are one of our main results.

\begin{figure}[!t]
\centering
\includegraphics[width = 0.5\textwidth]{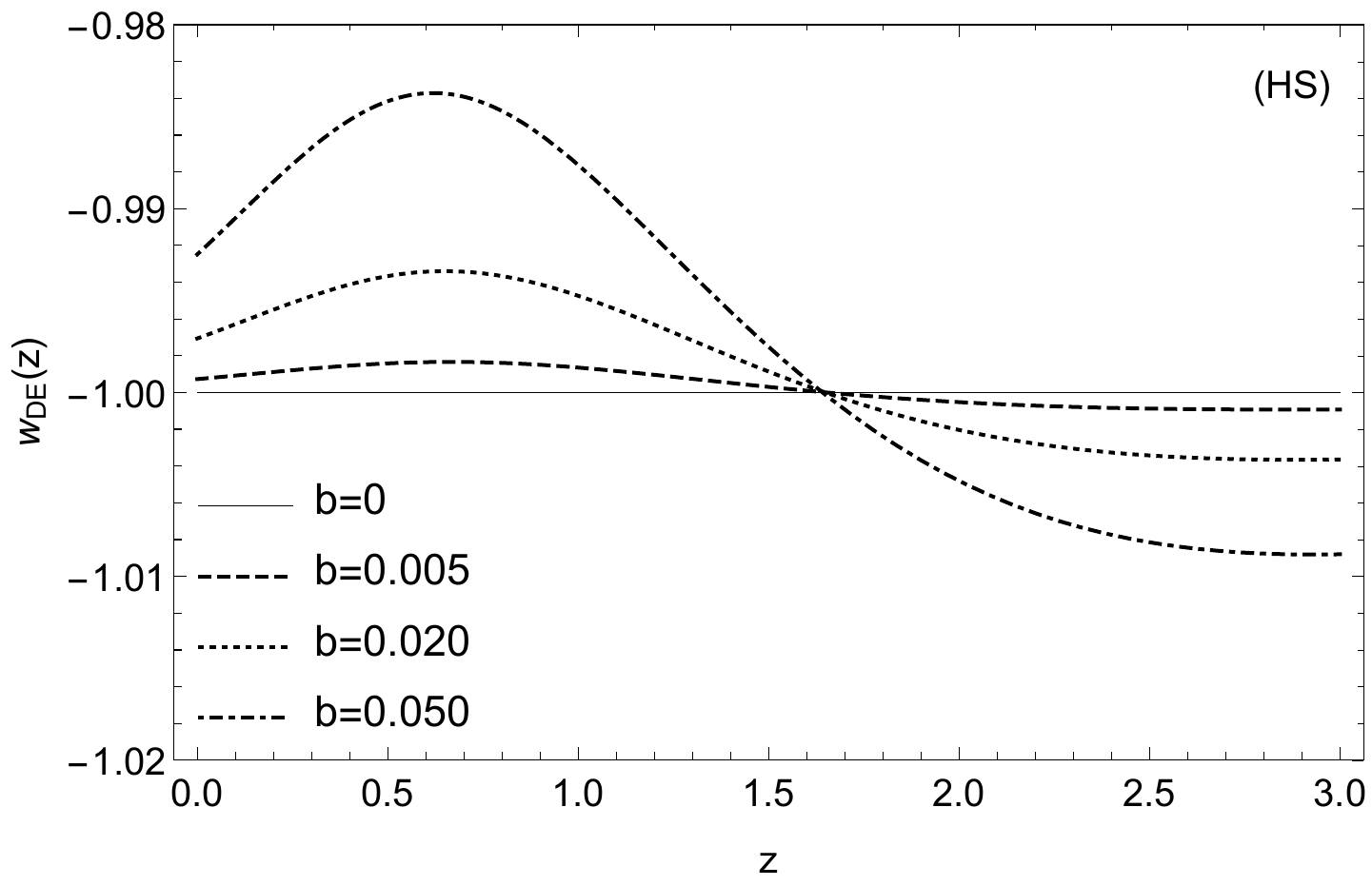}
\caption{The DE equation of state $w_{DE}(z)$ for the HS model for $\Omega_{m0}=0.3$, $n=1$ and for a variety of values of the parameter $b$, with $b \in [0,0.05]$. As can be seen, the equation of state crosses $w_{DE}=-1$ at approximately the same redshift $z\sim1.65$. At early times, we have $1+w_{DE}<0$ thus violating the SEC.}
\label{fig:wdeHS}
\end{figure}

Finally, for our effective DE fluid in Eq. \eqref{eq:effTmn} the most common energy conditions \cite{wald1984general} can be written in terms of the effective DE density and pressure:

\bea
{\bf NEC} &\Longrightarrow & \bar{\rho}_{DE}+\bar{P}_{DE} \ge 0, \nn \\
{\bf WEC} &\Longrightarrow &\bar{\rho}_{DE}\ge 0 \hspace{3mm}\text{and}\hspace{3mm} \bar{\rho}_{DE}+\bar{P}_{DE} \ge 0, \nn \\
{\bf DEC} &\Longrightarrow & \bar{\rho}_{DE}\ge 0 \hspace{3mm}\text{and}\hspace{3mm} \bar{\rho}_{DE} \ge \left|\bar{P}_{DE}\right|, \nn \\
{\bf SEC} &\Longrightarrow & \bar{\rho}_{DE}+3\bar{P}_{DE} \ge 0 \hspace{3mm}\text{and}\hspace{3mm} \bar{\rho}_{DE}+\bar{P}_{DE} \ge 0, \nn
\eea
where NEC, WEC, DEC and SEC correspond respectively to the null, weak, dominant and strong energy conditions. As expected for an accelerating universe \cite{santos2007energy,visser2000energy}, we have checked that the SEC is violated. Since the condition $\bar{\rho}_{DE} \ge 0$ holds, we find that the NEC, WEC and DEC can be translated into the following constraint for the DE equation of state $w_{DE} \ge -1$. As can be seen in Fig.~\ref{fig:wdeHS} for the HS model, the NEC, WEC and DEC are violated for redshifts $z \gtrsim 1.65$ for reasonable values of the parameter $b$ (see Eq.~\eqref{Hu1} in the next section), for $b \in [0,0.05]$.

\subsection{Results for specific f(R) models}
\label{subsection:specific-models-results}

So far, our analysis has been quite general and here we work out a couple of examples. In this section we will present our results for two specific models, namely, the HS model and the so-called designer (DES) model which has an expansion history equal to the \lcdm model. These models are interesting because they satisfy solar system tests and give a proper matter era. Note, however, that in the literature one finds other $f(R)$ models sharing these properties (see, for instance, Refs.\cite{Starobinsky:2007hu,Cognola:2007zu,Dunsby:2010wg}), but to simplify our presentation we only focus on the two aforementioned models.

Since modifications to GR are expected to become important at late times, we consider a universe only containing matter and an effective DE fluid.\footnote{In this paper we focus on the late-time evolution of the Universe, but it is possible that MG theories play a part in earlier stages as well, namely, the inflationary period. There exist $f(R)$ models that give a unified description of early- and late-time accelerating phases of the Universe \cite{Nojiri:2006gh,Elizalde:2010ts} and our effective fluid approach could in principle also be applied in these scenarios.} The system of differential equations that we are interested in is, hence, given by Eqs.~\eqref{eq:phiprimeeq}, \eqref{eq:anisoeq}, \eqref{Eq:evolution-delta}, \eqref{Eq:evolution-V}:
\bea
\delta_m' = 3 \Phi'-\frac{V_m}{a^2 H} , \label{Eq:system-ode-1}
\eea
\bea
V_m' = -\frac{V_m}{a}+\frac{k^2}{a^2 H}\Psi ,
\eea
\bea
\delta_{DE}' = 3(1+w_{DE}) \Phi'-\frac{V_{DE}}{a^2 H} & & \nn \\
-\frac{3}{a}\left(\frac{\delta P_{DE}}{\bar{\rho}_{DE}}-w_{DE}\delta_{DE}\right),
\eea
\bea
V_{DE}' = -(1-3w_{DE})\frac{V_{DE}}{a} + \frac{k^2}{a^2 H}\frac{\delta P_{DE}}{\bar{\rho}_{DE}} & & \nn \\
 + (1+w_{DE})\frac{k^2}{a^2 H} \Psi-\frac23 \frac{k^2}{a^2 H} \pi_{DE},
\eea
\bea
\frac{k^2}{a^2}\Phi+3 H^2(a \Phi'+\Psi)=-\frac32(\Omega_m \delta_m + \Omega_{DE} \delta_{DE}), & &
\eea
\bea
\frac{k^2}{a^2}(\Phi-\Psi)=3\Omega_{DE} \pi_{DE}, & & \label{Eq:system-ode-2}
\eea
where the prime $'$ denotes a derivative with respect to scale factor $a$, we have assumed that the matter component is cold ($w_m\simeq0$) and pressureless ($c_{s,m}^2\simeq0$), $\Omega_m=\Omega_{m0}a^{-3}$, $\Omega_{DE}=H^2-\Omega_m$, and finally that the effective DE density, pressure and velocity perturbations are given by Eqs.~(\ref{eq:effpres}),(\ref{eq:effder}) and (\ref{eq:efftheta}), respectively.

\subsubsection{The HS model}

The HS model \cite{Hu:2007nk} has a lagrangian\footnote{The Starobinsky model \cite{Starobinsky:2007hu} has a lagrangian $f(R)=R-c_1~m^2 \left[1-\left(1+R^2/m^{4}\right)^{-n}\right]$ and the results we obtain are very similar to those for the HS model. To keep our presentation simple we will only present results  for the HS model.} given by

\begin{equation}
\label{Hu}
f(R)=R-m^2 \frac{c_1 (R/m^2)^n}{1+c_2 (R/m^2)^n},
\end{equation}
where $c_1$, $c_2$ are two free parameters, $m^2\simeq \Omega_{m0}H^{2}_{0}$ is of the order of the Ricci scalar $R_{0}$, $H_{0}$ is the Hubble constant, $\Omega_{m0}$ is the dimensionless matter density today; and $m$ and $n$ are positive constants with $n$ usually taking positive integer values i.e., $n=1, 2, \cdots$. In the rest of our paper we assume $n=1$.

After simple algebraic manipulations Eq.~(\ref{Hu}) can also be written as \cite{Basilakos:2013nfa}
\bea
\label{Hu1}
f(R)&=& R- \frac{m^2 c_1}{c_2}+\frac{m^2 c_1/c_2}{1+c_2 (R/m^2)^n} \nn\\
&=& R- 2\Lambda\left(1-\frac{1}{1+(R/(b~\Lambda)^n}\right) \nn \\
&=& R- \frac{2\Lambda }{1+\left(\frac{b \Lambda }{R}\right)^n},
\eea
where $\Lambda= \frac{m^2 c_1}{2c_2}$ and $b=\frac{2 c_2^{1-1/n}}{c_1}$. In this form it is clear that this model can be arbitrarily close to $\Lambda$CDM, depending on the parameters $b$ and $n$. Moreover, for $n>0$ it has the limits \cite{Basilakos:2013nfa}:
\bea
\lim_{b\rightarrow0}f(R)&=&R-2\Lambda , \nn \\
\lim_{b\rightarrow \infty}f(R)&=&R.
\eea
Since the HS model tends to \lcdm for $b\rightarrow 0$, it can be considered as a small perturbation around the \lcdm model. Therefore, it should come as no surprise that the HS model can successfully pass the solar system tests.

Furthermore, in Ref.~\cite{Basilakos:2013nfa} it was shown that for small values of the parameter $b$ one is always able to find an analytic approximation to the Hubble parameter that works to a level of accuracy better than $\sim10^{-5}\%$ when the parameter $b$ is of the order of $b\sim [0.001-0.1]$, thus making the approximations very useful. Then, the Hubble parameter $H(t)=\frac{da/dt}{a}$ can be well approximated by
\be
H_{HS}(a)^2=H_\Lambda(a)^2+b~\delta H_1(a)^2+b^2~\delta H_2(a)^2+ \cdots, \label{eq:HubHSapp}
\ee
where the functions $\delta H_1(a)$ and $\delta H_2(a)$ are given in the Appendix of \cite{Basilakos:2013nfa}.

From Eqs. \eqref{eq:wde},\eqref{Hu1},\eqref{eq:HubHSapp} and considering a universe only containing matter and DE, we can calculate the DE equation of state as a series expansion in terms of $b$
\begin{widetext}
\be
w_{DE}(a)\simeq -1-\frac{12 \left(a^3 (\Omega_{m0}-1) \Omega_{m0} \left(a^3 (\Omega_{m0}-1)-\Omega_{m0}\right) \left(8 a^3 (\Omega_{m0}-1)+\Omega_{m0}\right)\right)}{\left(\Omega_{m0}-4 a^3 (\Omega_{m0}-1)\right)^4}b+\cdots,\label{eq:wdeHSapp}
\ee
\end{widetext}
while the DE anisotropic stress will be given by
\bea
\pi_{DE}(a)&=& \frac{1}{F}\frac{\frac{k^2}{a^2}\frac{F_{,R}}{F}}{1+3\frac{k^2}{a^2}\frac{F_{,R}}{F}}\frac{\bar{\rho}_m}{\bar{\rho}_{DE}} \delta_m \nn \\
&\simeq& \left(\frac{k^2}{a^2}\frac1{H_0^2}\frac{4 a^9 (1-\Omega_{m0})^2}{3 \left(\Omega_{m0} +4 a^3 (1-\Omega_{m0} )\right)^3}b+\cdots\right)\times \nn \\
 & & \frac{\bar{\rho}_m}{\bar{\rho}_{DE}} \delta_m.
\eea

From the system of differential equations \eqref{Eq:system-ode-1}-\eqref{Eq:system-ode-2} and the DE perturbations \eqref{eq:effpres}-\eqref{eq:effpi} we can derive approximate solutions in a matter dominated regime ($H(a)^2/H_0^2\simeq \Omega_{m0} a^{-3}$):
\bea
w_{DE}(a)&\simeq& -1-\frac{12 a^3 b (1-\Omega_{m0})}{\Omega_{m0}}+\cdots,
\eea
\bea
\frac{\delta P_{DE}(a)}{\bar{\rho}_{DE}(a)}&\simeq& b (1-\Omega_{m0})^2\left(\frac{8 a^7 k^2}{9 \Omega_{m0}^3 H_0^2}-\frac{66 a^5 H_0^2}{k^2 \Omega_{m0}}+\cdots\right)\times \nn \\
 & & \frac{\Omega_m(a)}{\Omega_{DE}(a)} \delta_m,
\eea
\bea
\pi_{DE}&\simeq& b\left(\frac{4 a^7 k^2 (1-\Omega_{m0})^2}{3 \Omega_{m0}^3 H_0^2}+\cdots\right) \frac{\Omega_m(a)}{\Omega_{DE}(a)} \delta_m,
\eea
\bea
\delta_m(a)&\simeq& \delta_0 \left(a+\frac{3 \Omega_{m0}H_0^2}{k^2}\right),
\eea
\bea
V_m(a)&\simeq& -\delta_0\sqrt{a \Omega_{m0}}+\cdots,
\eea
\bea
\delta_{DE}(a)&\simeq& -\delta_0 \; b \; (1-\Omega_{m0})\left(\frac{a^5 k^2}{3 \Omega_{m0}^2 H_0^2} + \right. \nn \\
& & \left. {} \frac{8 a^4}{35 \Omega_{m0}}-\frac{495 a^3 H_0^2}{13 k^2}-\frac{594 a^2 H_0^4\Omega_{m0}}{5 k^4}+\cdots\right), \nn \\
& &
\eea
\bea
V_{DE}(a)&\simeq& \delta_0 \; b \; (1-\Omega_{m0})\left(-\frac{396\;a^{5/2} H_0^2\sqrt{\Omega_{m0}}}{13 k^2}- \right. \nn \\
& & \left. {} \frac{32 a^{7/2}}{5 \sqrt{\Omega_{m0}}}+\cdots\right),
\eea
\bea
\Phi(a)&\simeq& -\frac32 \delta_0 \frac{\Omega_{m0}H_0^2}{k^2}+\cdots ,
\eea
where $\Omega_m(a)=\Omega_{m0}a^{-3}$ and in this limit $\Omega_{DE}(a)\simeq 1-\Omega_{m0}$. Also, as can be seen from the above expressions, the dominant contributions in the sub-horizon limit and in the matter-dominated regime are $\delta_{DE}\propto k^2 a^5$ and $V_{DE}\propto a^{7/2}$. When numerically solving the system of differential equations \eqref{Eq:system-ode-1}-\eqref{Eq:system-ode-2}, we will use the above solutions as initial conditions.

\subsubsection{The DES model}

The DES model \cite{Multamaki:2005zs,delaCruzDombriz:2006fj,Nesseris:2013fca}, which has a background exactly that of the \lcdm model, has a lagrangian given by
\bea
\label{des}
f(R)&=&R-2\Lambda+\alpha~H_0^2\left(\frac{\Lambda }{R-3 \Lambda }\right)^{c_{0}} \times \nn \\
 & & {}_2F_1\left(c_{0},\frac{3}{2}+c_{0},\frac{13}{6}+2c_{0},\frac{\Lambda }{R-3 \Lambda }\right)\;,
\eea
where $c_{0}=\frac{1}{12} \left(-7+\sqrt{73}\right)$ and $\alpha$ is a free dimensionless parameter.

While for the DES model the background is much simpler than for the HS model (in the DES model the expansion history matches that of the \lcdm model, i.e., $H_{DES}^2(a)=H_{\Lambda CDM}^2(a)$, Eq.~\eqref{des} makes more complicated the expressions for all the effective DE quantities. We have found an approximation around $a\simeq 0$ that works very well in the range $a\in[0,1]$; it reads
\bea
&F(a)&\simeq 1 + \nn \\
& & f_{R,0}\frac{\Omega_{m0}^{-c_0-1}}{\, _2F_1\left(c_0+1,c_0+\frac{3}{2};2 c_0+\frac{13}{6};1-\Omega_{m0}\right)}a^{3 (1 + c_0)} \nn \\
&+&\mathcal{O}(a^{3(2+c_0)}) , \label{eq:designF}
\eea
where $f_{R,0}\equiv F(a=1)-1$. For viable models, the parameter $f_{R,0}$ has typical values on the order $f_{R,0}\sim -10^{-4}$ (see, for instance, Ref.  \cite{Pogosian:2007sw}).\footnote{For illustration purposes we note that the right-hand side of Eq.~(\ref{eq:designF}) evolves roughly as
$F(a)\thickapprox 1+0.85\;f_{R,0}\;\Omega_{m0}^{-0.57} a^{3.386}$.
We however do not use this expression in our computations.}

Following the same approach as for the HS model we have found approximate solutions in a matter dominated regime
\bea
w_{DE}(a)&=& -1,
\eea
\bea
\frac{\delta P_{DE}}{\bar{\rho}_{DE}}&\simeq& \left(-\frac{2 (c_0+1) f_{R,0} k^2 a^{3 c_0+4} \Omega_{m0}^{-c_0-2}}{9 \; g_0}+\cdots\right) \times \nn \\
& & \frac{\Omega_m(a)}{\Omega_{DE}(a)} \delta_m,
\eea
\bea
\pi_{DE}&\simeq& \left(-\frac{(c_0+1) f_{R,0} k^2 a^{3 c_0+4} \Omega_{m0}^{-c_0-2}}{3 \; g_0}+\cdots\right) \times \nn \\
& & \frac{\Omega_m(a)}{\Omega_{DE}(a)} \delta_m,
\eea
\bea
\delta_m(a)&\simeq& \delta_0 \left(a+\frac{3 \Omega_{m0}H_0^2}{k^2}\right),
\eea
\bea
V_m(a)&\simeq& -\delta_0\sqrt{a \Omega_{m0}}+\cdots,
\eea
\bea
\delta_{DE}(a)&\simeq& \frac{\delta_0 f_{R,0} a^{1+3c_0} \Omega_{m0}^{-1-c_0} \left(a (1+2c_0) k^2+36 c_0 \Omega_{m0}\right)}{9 \; g_0 \; (1-\Omega_{m0})} \nn \\
 &+& \cdots ,
\eea
\bea
V_{DE}(a)&\simeq& 0+\cdots,
\eea
\bea
\Phi(a)&\simeq& -\frac32 \delta_0 \frac{\Omega_{m0}H_0^2}{k^2}+\cdots,
\eea
where $g_0={}_2F_1\left(1+c_0,\frac32+c_0,\frac{13}{6}+2c_0,1-\Omega_{m0}\right)$. In the next section, we will use these approximations as initial conditions for the numerical evolution in the effective fluid approach.

Note that in Ref.~\cite{Sapone:2009mb} the authors derived approximations to the evolution of the DE density contrast $\delta_{DE}\simeq \delta_0 (1+w) \left(\frac{a}{1-3w}+\frac{3 H0^2 \Omega_{m0}}{k^2} \right)$ and velocity perturbation $V_{DE}\simeq-\delta_0 (1+w) H_0 \sqrt{\Omega_{m0}}a^{1/2}$. Clearly, in both cases when $w=-1$, as is the case for the DES model, we would have that $(\delta_{DE},V_{DE})=(0,0)$ as expected. However, we have seen that the DE perturbations in the DES model (despite having $w_{DE}=-1$) have in general a dependence on the scale factor $a$ which is quite different. Therefore, care should be used when applying the expressions of Ref.~\cite{Sapone:2009mb} as initial conditions and instead one should derive again the correct expressions as we have done. 

\section{Numerical solution of the evolution equations}
\label{Section:numerical-solution}

\subsection{Evolution of perturbations}

Here we present the results of the numerical solution of the evolution equations \eqref{Eq:system-ode-1}-\eqref{Eq:system-ode-2}. In all cases we will assume $\Omega_{m0}=0.3$, $k=300H_0$, $f_{R0}=-10^{-4}$ and $\sigma_{8,0}=0.8$, where $f_{R,0}=F(a=1)-1$, unless otherwise specified. We set the initial conditions well inside the matter dominated regime at $a=10^{-3}$. The reason we choose the specific value of $k=300H_0\sim 0.1\;h/\textrm{Mpc}$ for the wave-number is that it corresponds to the largest value of $k$ we can choose without entering the non-linear regime.

\begin{figure*}[!t]
\centering
\includegraphics[width = 0.495\textwidth]{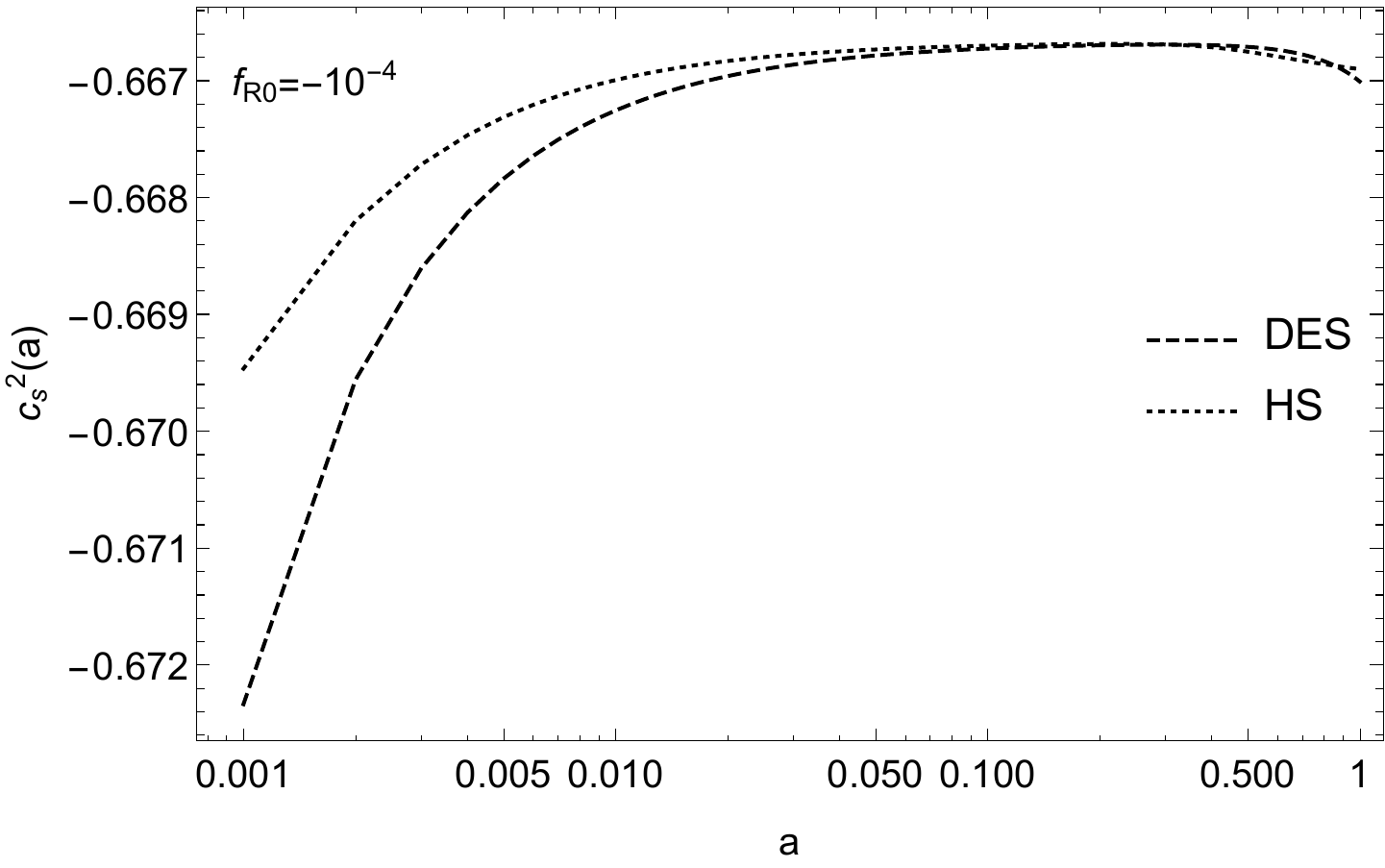}
\includegraphics[width = 0.495\textwidth]{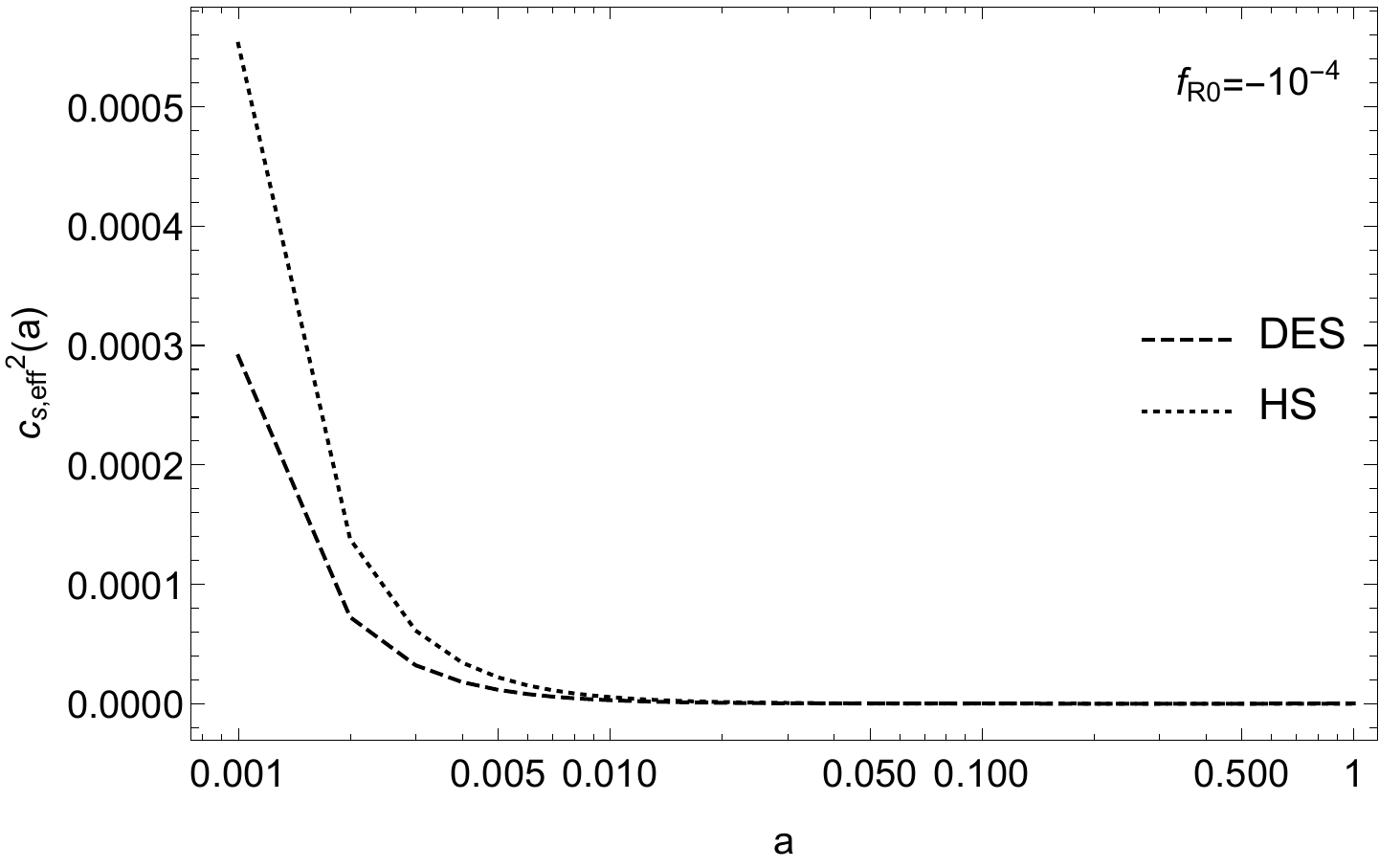}
\caption{The DE fluid sound speed $c_{s,DE}^2$ (left) and the DE effective sound speed $c_{s,eff}^2$ (right) given by Eqs.~(\ref{eq:cs2de}) and (\ref{eq:cs21}) for both the HS (dotted line) and DES (dashed lines) models for $\Omega_{m0}=0.3$, $k=300H_0$ and $f_{R,0}=-10^{-4}$. As can be seen, for both models the DE sound speed remains close to $c_{s,DE}^2\sim-\frac23$ while the DE effective sound speed is close to $c_{s,eff}^2\sim 0^{+}$. }
\label{fig:soundspeed}
\end{figure*}

Before we proceed with the discussion of our results, it is instructive to show the evolution of the DE sound speed $c_{s,DE}^2$ and the DE effective sound speed $c_{s,eff}^2$ given by Eqs.~(\ref{eq:cs2de}) and (\ref{eq:cs21}), respectively, for both the HS and DES models. The plots are shown in Fig.~\ref{fig:soundspeed}, where we show $c_{s,DE}^2$ (left) and $c_{s,eff}^2$ (right) for both the HS (dotted line) and DES (dashed lines) models. As can be seen, for both models the DE sound speed remains close to $c_{s,DE}^2\sim-\frac23$ while the effective sound speed is close to $c_{s,eff}^2\sim 0^{+}$. On the one hand, this behavior implies that at early times while the DE effective sound speed is positive, the DE perturbations are expected to grow. On the other hand, at late times as the DE effective sound speed goes to zero asymptotically the DE perturbations are expected to reach a plateau and stop growing.

\begin{figure*}[!t]
\centering
\includegraphics[width = 0.495\textwidth]{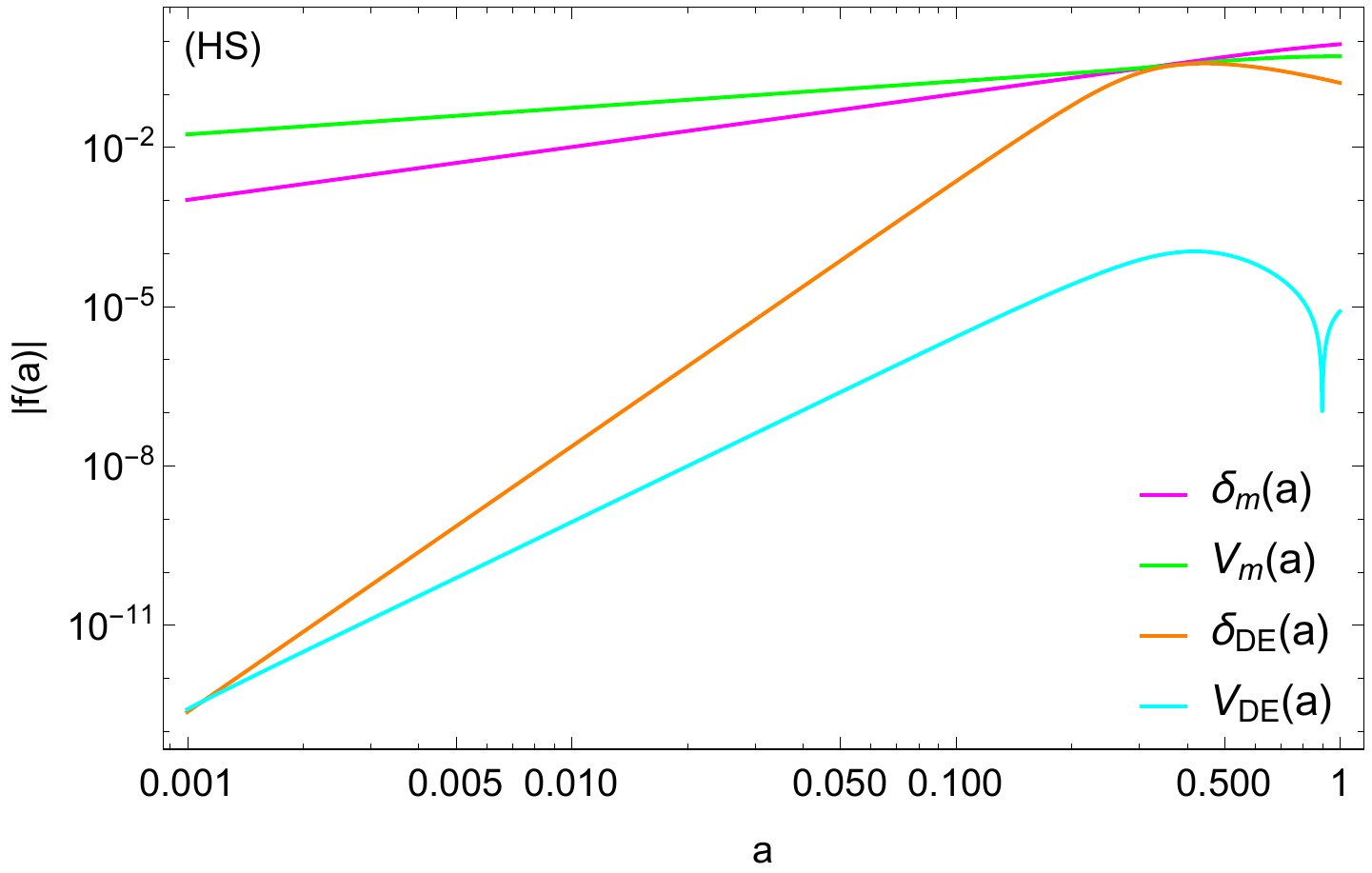}
\includegraphics[width = 0.495\textwidth]{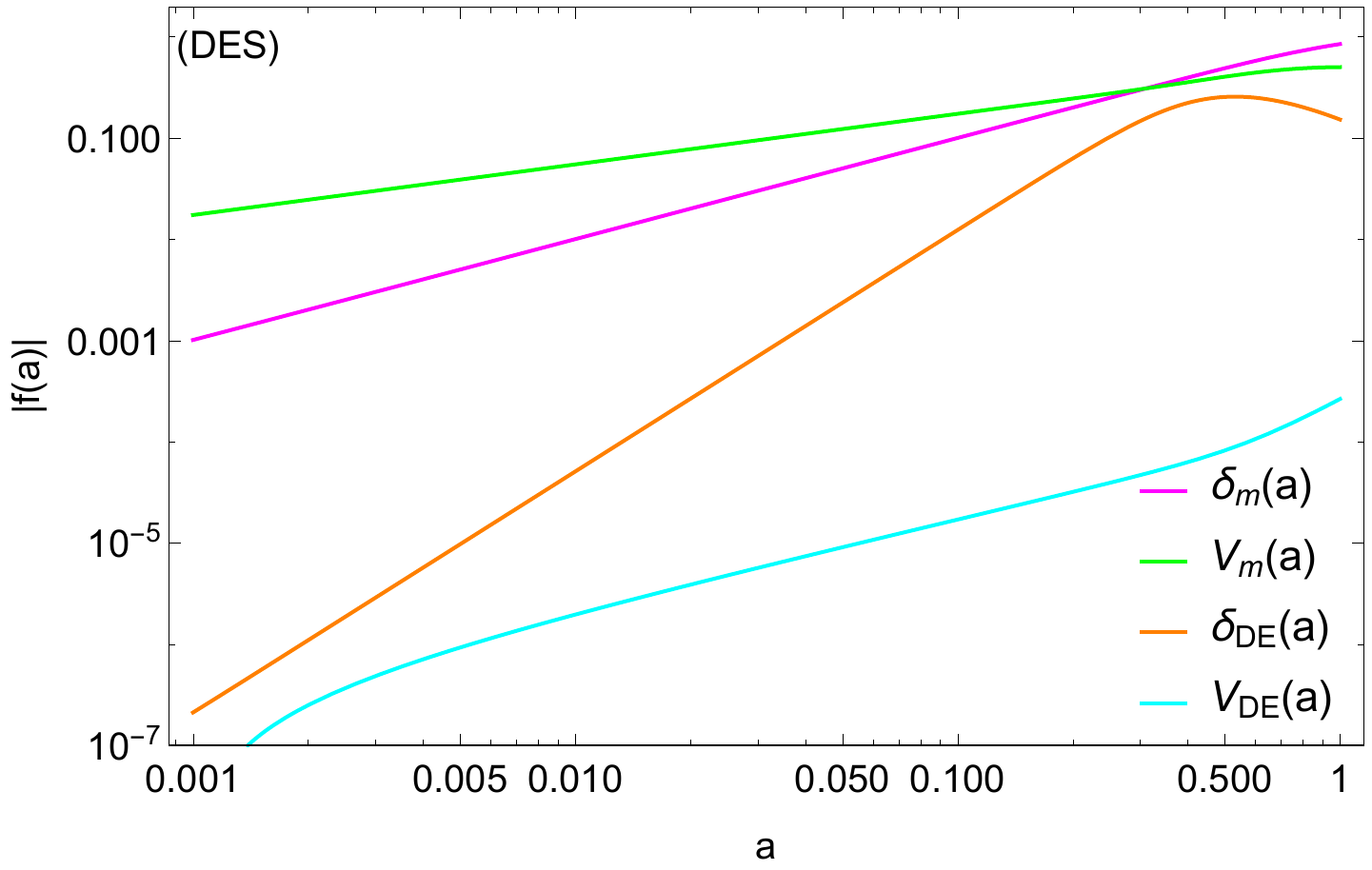}
\caption{The evolution of the matter and effective DE perturbation variables $(\delta_m,V_m,\delta_{DE},V_{DE})$ for the HS (left) and the DES (right) models  for $\Omega_{m0}=0.3$, $k=300H_0$, $\delta_0=1$, and $f_{R,0}=-10^{-4}$. As described in the text, the DE perturbations reach a plateau and then flatten out for both models, as expected from the fact that the DE effective sound speed given by Eq.~(\ref{eq:cs21}) goes to zero at late times (see Fig.~\ref{fig:soundspeed}). Also, in all cases, the DE velocity perturbation remains significantly suppressed with respect to the rest of the variables.}
\label{fig:perturb}
\end{figure*}

\begin{figure*}[!t]
\centering
\includegraphics[width = 0.495\textwidth]{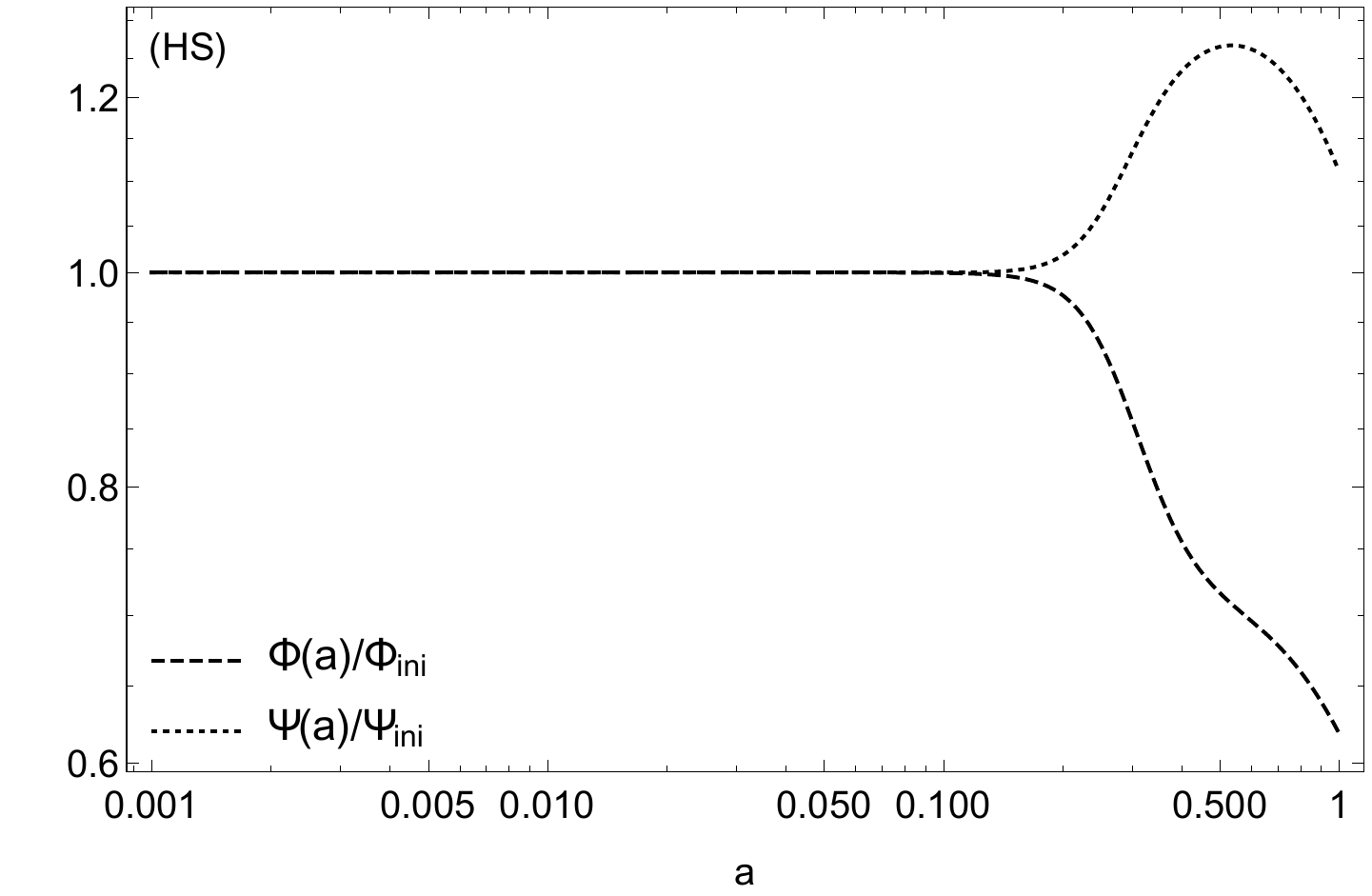}
\includegraphics[width = 0.495\textwidth]{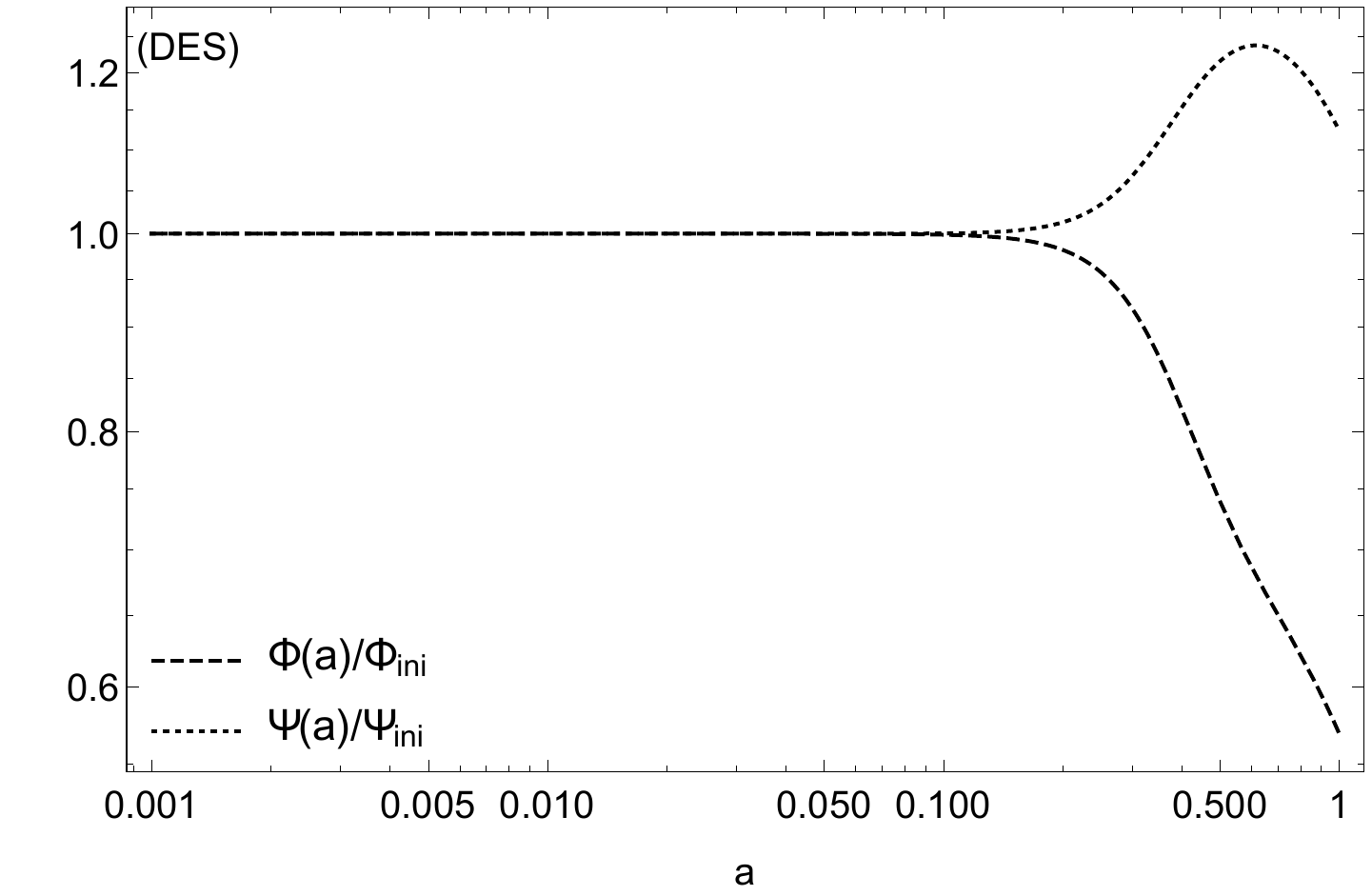}
\caption{The evolution of the potentials $\Phi$ and $\Psi$ for the HS (left) and the DES (right) models for $\Omega_{m0}=0.3$, $k=300H_0$, $\delta_0=1$, and $f_{R,0}=-10^{-4}$. Due to a non-vanishing DE anisotropic stress the potentials diverge from each other at late times.}
\label{fig:potentials}
\end{figure*}

In Figs.~\ref{fig:perturb} and \ref{fig:potentials} we present our results for the perturbation variables $(\delta_m,V_m,\delta_{DE},V_{DE})$ and the potentials $(\Phi, \Psi)$, respectively. 
As noted before, the DE perturbations reach a plateau and then flatten out for both models, as expected from the fact that the DE effective sound speed goes to zero at late times (see Fig.~\ref{fig:soundspeed}). Also, in all cases, the DE velocity perturbation remains significantly suppressed with respect to the rest of the variables. Furthermore, the potentials remain approximately equal until $a\sim 0.1$, which as seen in Fig.~\ref{fig:soundspeed} corresponds to the epoch when roughly $c_{s,eff}^2\sim 0$, and then diverge from each other significantly due to the presence of the anisotropic stress.

\subsection{Growth rate of matter perturbations}

\begin{figure*}[!t]
\centering
\includegraphics[width = 0.495\textwidth]{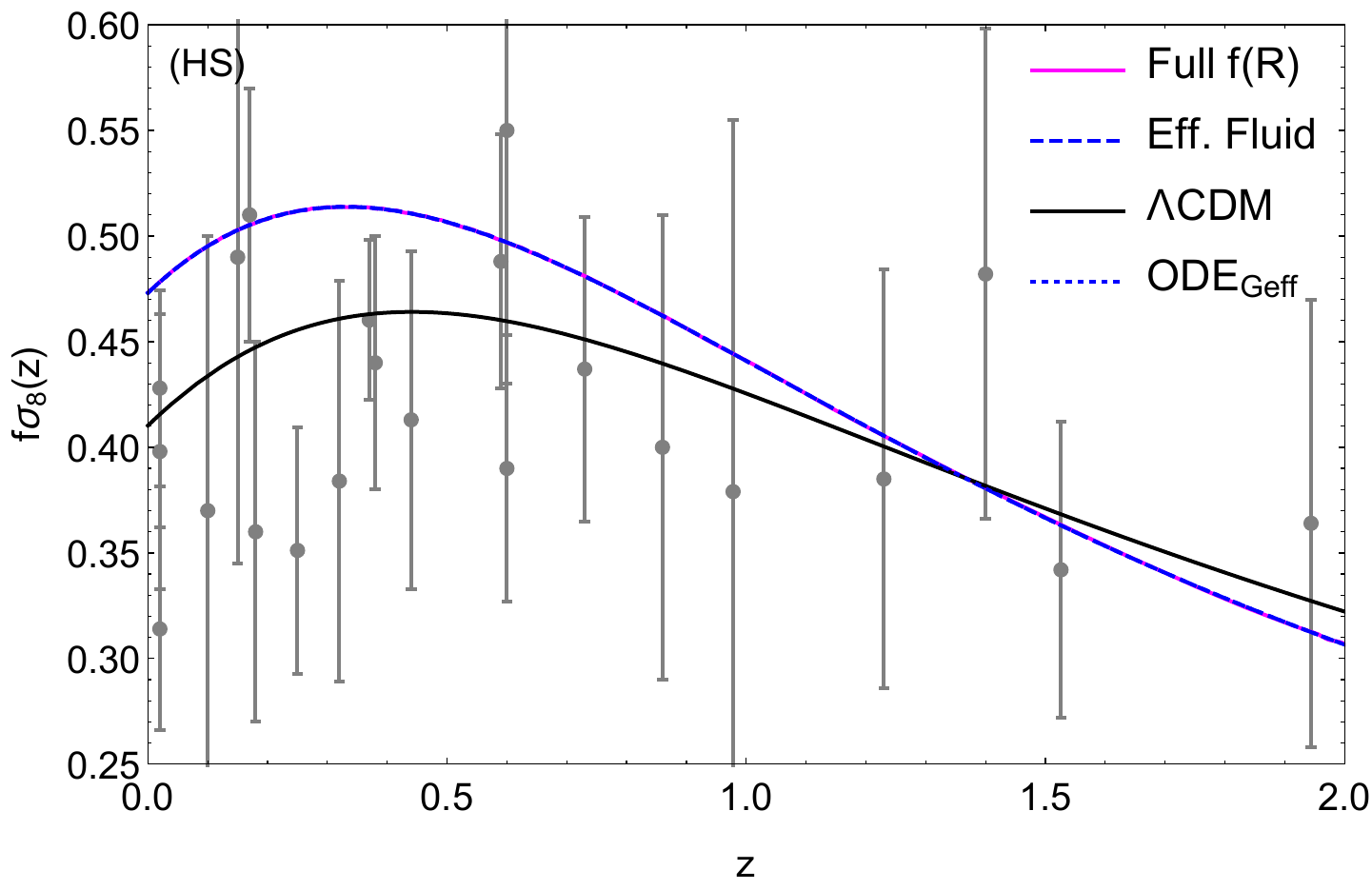}
\includegraphics[width = 0.495\textwidth]{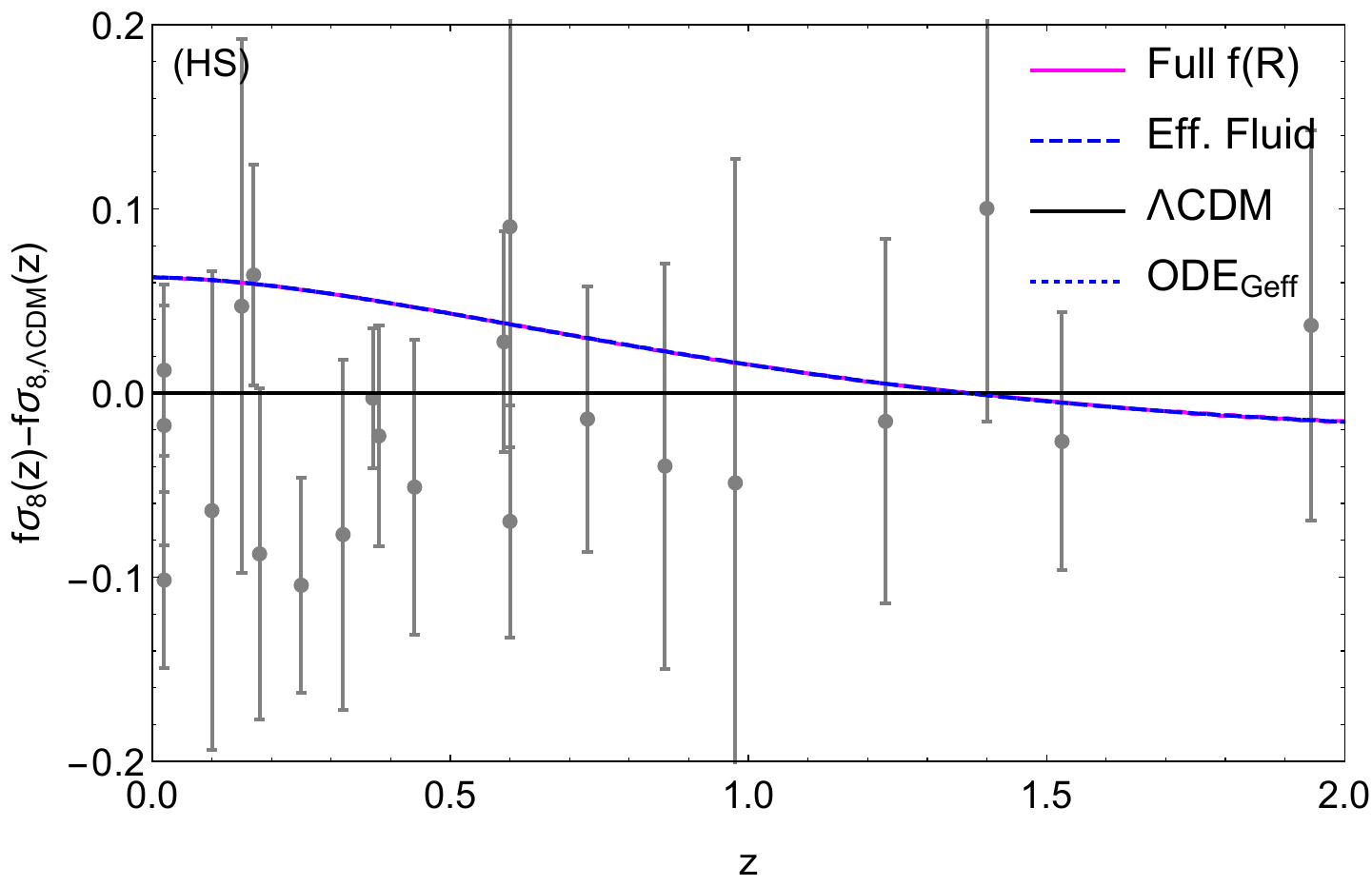}
\caption{The evolution of the $f\sigma_8(z)$ parameter for the HS model for $\Omega_{m0}=0.3$, $k=300H_0$, $f_{R,0}=-10^{-4}$ and $\sigma_{8,0}=0.8$ versus the $f\sigma_8$ data compilation from Ref.~\cite{Sagredo:2018ahx}. On the left panel we show the theoretical curves for the ``Full $f(R)$" brute-force solution based on Ref.~\cite{Pogosian:2007sw} (magenta line), our effective fluid approach which we call ``Eff. Fluid" (blue dashed line), the $\Lambda$CDM model (black line) and the numerical solution of Eq. (\ref{eq:Geffode}) dubbed ``$\rm{ODE_{Geff}}$'' (dotted blue line). On the right panel we show the difference of the aforementioned theoretical curves with respect to that of the $\Lambda$CDM model. As can be seen, the agreement with all approaches is excellent.}
\label{fig:fs8HS}
\end{figure*}

Next we will also present our results for the growth rate of matter perturbations parameter $f\sigma_8(a)\equiv f(a)\cdot \sigma(a)$, where $f(a)=\frac{d ln\delta}{d lna}$ is the growth rate and $\sigma(a)=\sigma_{8,0}\frac{\delta(a)}{\delta(1)}$  is  the redshift-dependent root mean square (rms) fluctuations of the linear density field within spheres of radius $R=8 h^{-1} \textrm{\textrm{Mpc}}$, while the parameter $\sigma_{8,0}$ is its value today. This parameter is important as it can be shown to be not only independent of the bias $b_0$, but also a good discriminator of DE models \cite{Song:2008qt}.

In this section we will also compare our results with those of Ref.~\cite{Pogosian:2007sw} that follow a direct brute-force solution of the differential equations of the $f(R)$ model, dubbed ``Full $f(R)$" from now on. There is of course also the equation of state approach of Ref.~\cite{Battye:2015hza} and we have explicitly checked that our results are in excellent agreement with it; thus, to avoid an overload in both the presentation and the plots, in what follows we will only present the comparison with the ``Full $f(R)$'' approach. 

Both aforementioned approaches are exact, in the sense of having no approximations, however the one of Ref.~\cite{Pogosian:2007sw} suffers from the problem that the relevant equations are extremely stiff numerically, while in the one of Ref.~\cite{Battye:2015hza} the fluid equations are written in terms of a gauge-invariant entropy perturbation which cannot be easily translated to simple analytic expressions for the effective pressure, density contrast and velocity perturbations such as Eqs.~(\ref{eq:effpres}), (\ref{eq:effder}) and (\ref{eq:efftheta}) presented here.

\begin{figure*}[!t]
\centering
\includegraphics[width = 0.495\textwidth]{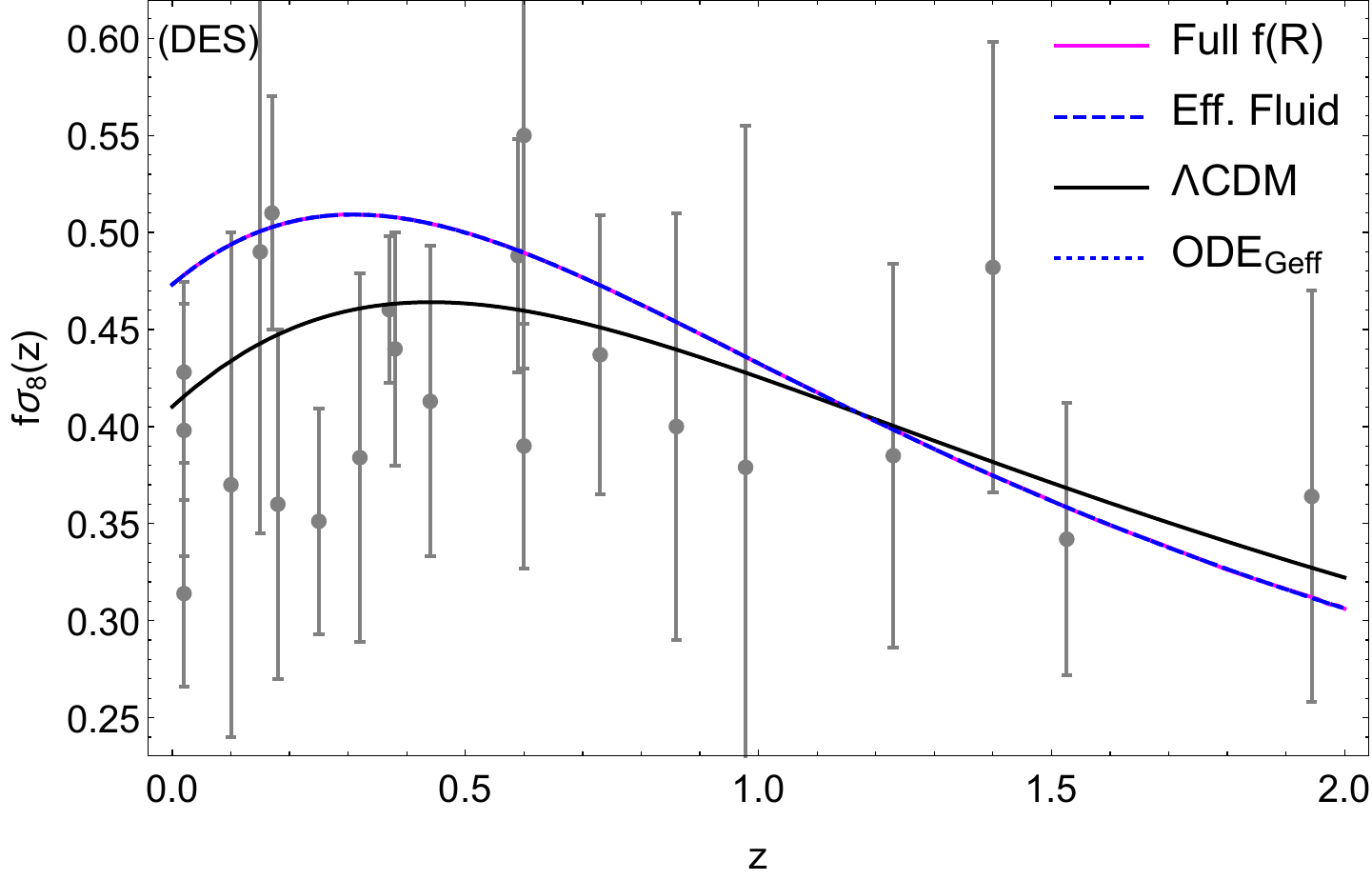}
\includegraphics[width = 0.495\textwidth]{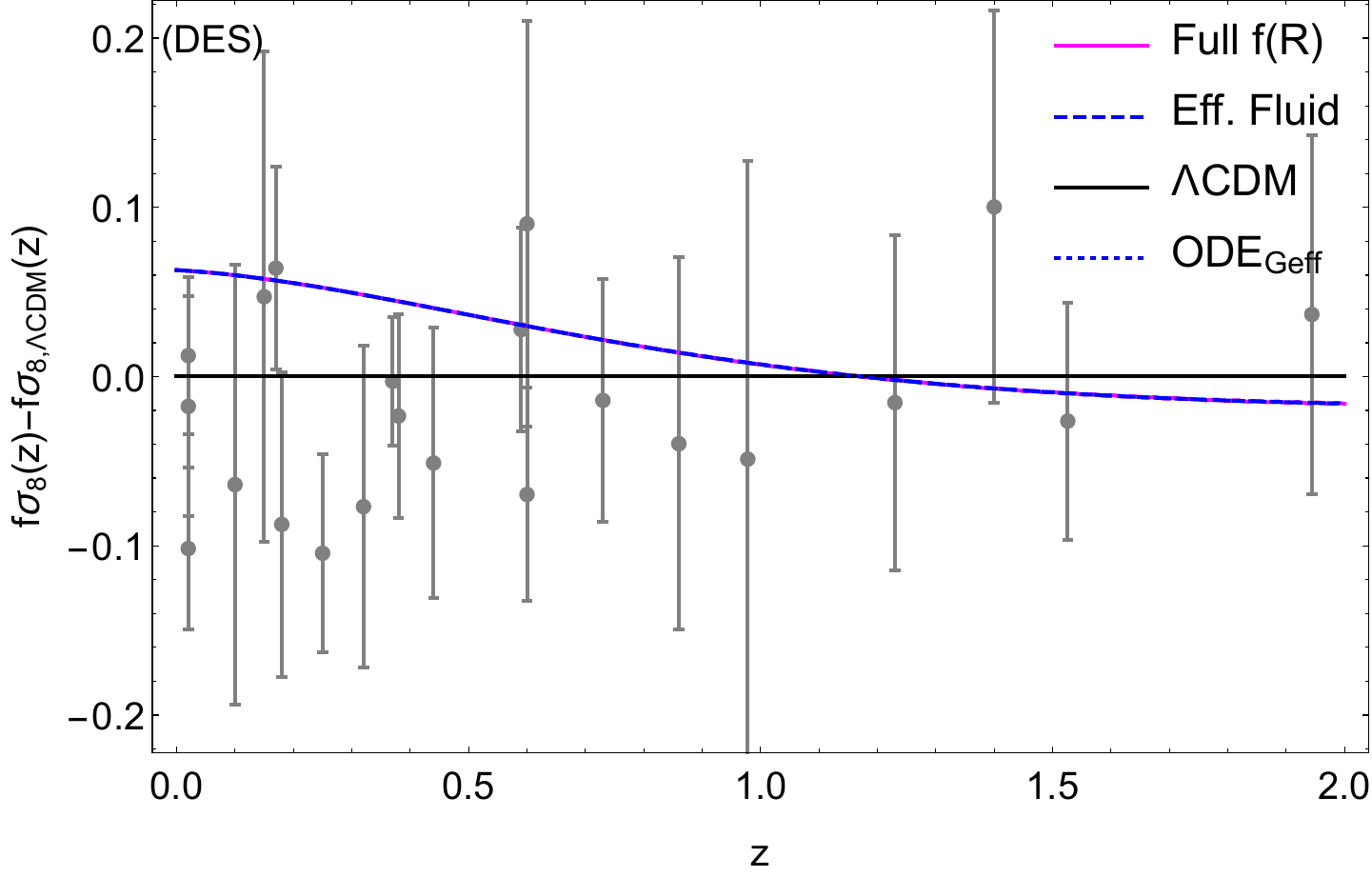}
\caption{The evolution of the $f\sigma_8(z)$ parameter for the DES model for $\Omega_{m0}=0.3$, $k=300H_0$, $f_{R,0}=-10^{-4}$ and $\sigma_{8,0}=0.8$ versus the $f\sigma_8$ data compilation from Ref.~\cite{Sagredo:2018ahx}. On the left panel we show the theoretical curves for the ``Full $f(R)$" brute-force solution based on Ref.~\cite{Pogosian:2007sw} (magenta line), our effective fluid approach which we call ``Eff. Fluid" (blue dashed line), the $\Lambda$CDM model (black line) and the numerical solution of Eq. (\ref{eq:Geffode}) dubbed ``$\rm{ODE_{Geff}}$'' (dotted blue line). On the right panel we show the difference of the aforementioned theoretical curves with respect to that of the $\Lambda$CDM model. As can be seen, the agreement with all approaches is excellent.}
\label{fig:fs8DES}
\end{figure*}

In Figs.~\ref{fig:fs8HS} and \ref{fig:fs8DES} we show the evolution of the $f\sigma_8(z)$ parameter for the HS and DES models respectively, for $\Omega_{m0}=0.3$, $k=300H_0$, $f_{R,0}=-10^{-4}$ and $\sigma_{8,0}=0.8$ versus the $f\sigma_8$ data compilation from Ref.~\cite{Sagredo:2018ahx}. On the left panel we show the theoretical curves for the ``Full $f(R)$" brute-force solution based on Ref.~\cite{Pogosian:2007sw} (magenta line), our effective fluid approach which we call ``Eff. Fluid" (blue dashed line), the $\Lambda$CDM model (black line) and the numerical solution of Eq. (\ref{eq:Geffode}) dubbed ``$\rm{ODE_{Geff}}$'' (dotted blue line). On the right panel we show the difference of the aforementioned theoretical curves with respect to that of the $\Lambda$CDM model. As can be seen, the agreement with all approaches is excellent.

\begin{figure*}[!t]
\centering
\includegraphics[width = 0.495\textwidth]{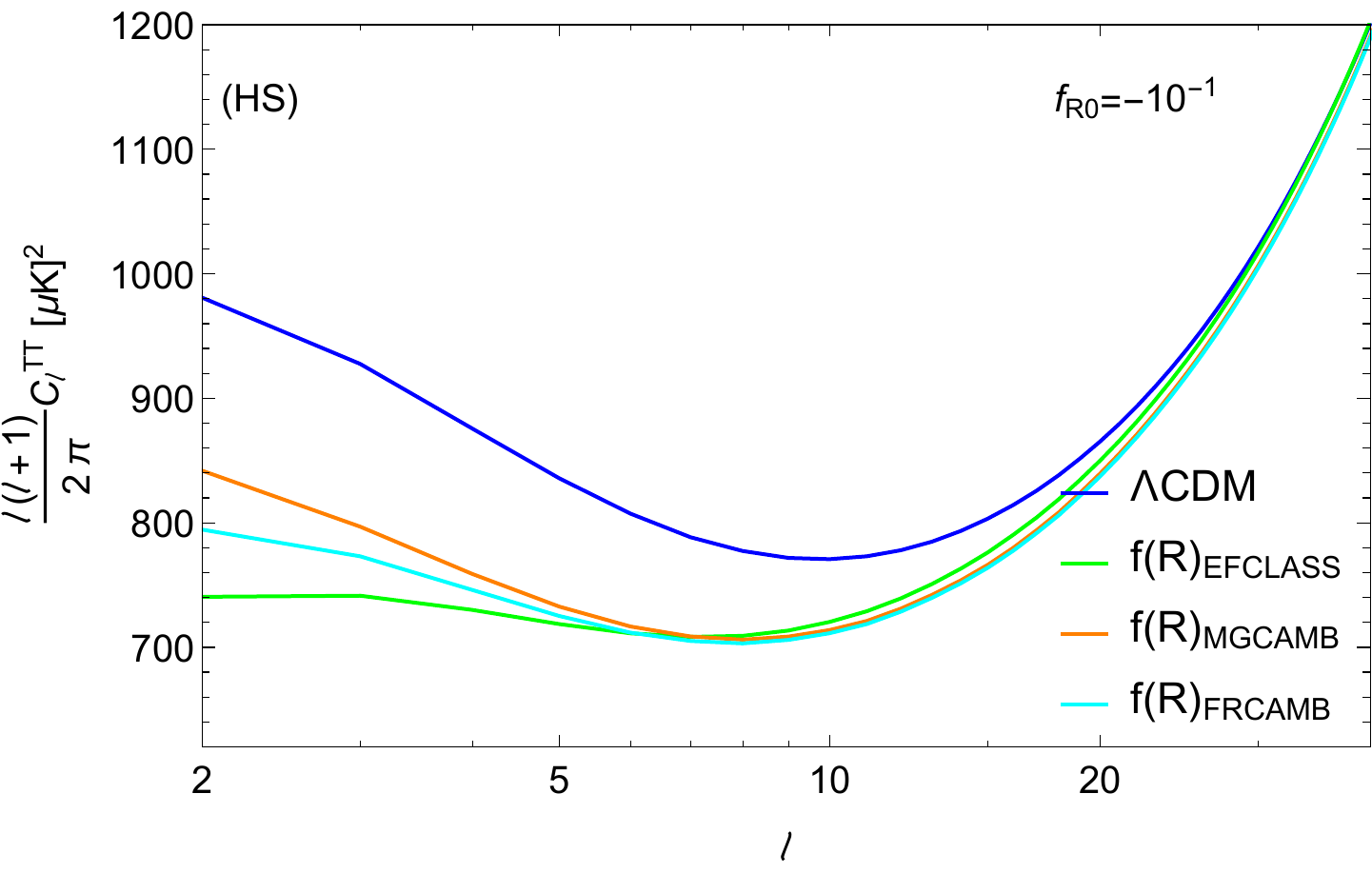}
\includegraphics[width = 0.495\textwidth]{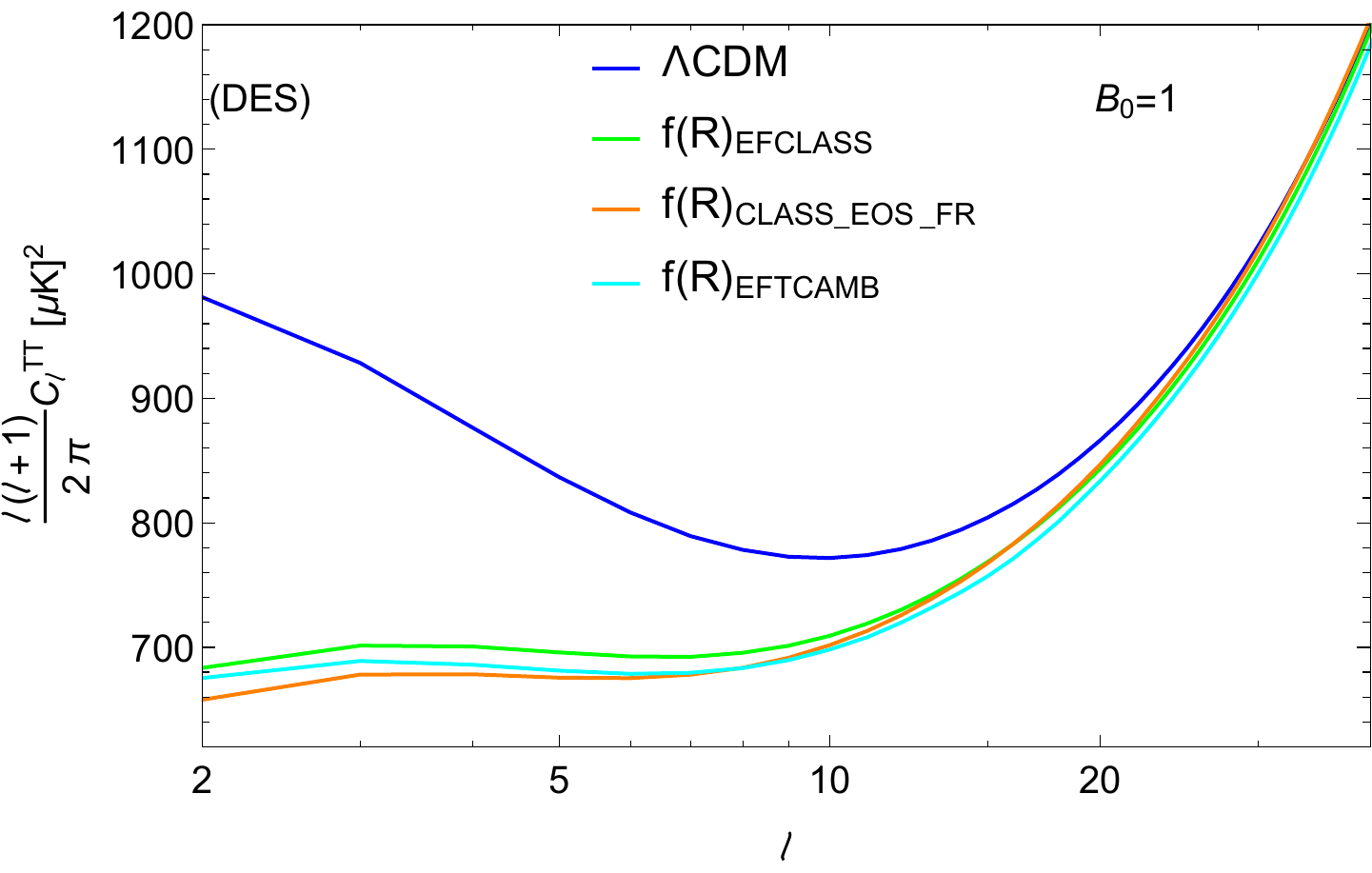}
\caption{The low multipoles of the unlensed CMB TT power spectrum for the HS model (left panel) and the DES model (right panel). We compare several codes: our own modifications to CLASS which we call EFCLASS, the codes MGCAMB and FRCAMB for the HS model and the codes CLASS\_EOS\_FR and EFTCAMB for the DES model. We find that in the case of the DES model all approaches are in very good agreement, but in the case of the HS model, which also requires modifying the background evolution, there is significant disagreement at $l\in[2,5]$ as the codes MGCAMB and FRCAMB do not take into account the change of the background properly. For these plots we assume $(n_s,A_s)=(1,2.3 \times 10^{-9})$, $f_{R,0}=-10^{-1}$ for the HS model and $B_0=1$, which corresponds to $f_{R,0}\simeq -0.159285$ for the DES model for $\Omega_{m0}=0.3$, while the rest of the parameters are as in the previous plots.}
\label{fig:cmbcls}
\end{figure*}

\subsection{CMB power spectrum}

We now also present the results for the CMB power spectra for both models and we compare our predictions with those of several other codes. As we show in Appendix \ref{Section:class-implementation}, our implementation of the effective fluid approach in the CLASS code \cite{Blas:2011rf}, while much simpler, also gives results in excellent agreement with other codes, such as EFTCAMB \cite{Hu:2013twa}, MGCAMB \cite{Zhao:2008bn}, FRCAMB \cite{He:2012wq}, CLASS\_EOS\_FR \cite{Battye:2017ysh}. In all cases, we took extreme care in order to match the various cosmological parameters between the codes and we explicitly tested that in the limit of the $\Lambda$CDM model, all codes agree with each other within the numerical errors. The fact that our implementation is consistent with that of Ref. \cite{Battye:2017ysh}, which is exact, shows the sub-horizon approximation can be safely applied in the models we discussed (see Fig.~\ref{fig:cmbcls}). This agrees with results in Ref. \cite{delaCruzDombriz:2008cp}: for $f(R)$ models that predict an accelerated expansion of the Universe and satisfy the local gravity constraints, the sub-horizon approximation is accurate.

In order to check with other results for the DES model in the literature, we find it advantageous to introduce the $B_0$ parameter defined as
\be
B_0=\frac{F_{,R}}{F}\frac{R'(a)}{a H'(a)/H(a)}|_{a=1}.
\ee
The main reason for this choice is that the effects of the modified gravity models on the ISW would be small for $f_{R,0}=-10^{-4}$ that we used in the previous plots. Thus in order to make the effect more visible and still be able to compare with other analyses, we will choose the value $B_0=1$, which corresponds to $f_{R,0}\simeq -0.159285$ for the DES model for $\Omega_{m0}=0.3$. For the HS model we will use $f_{R,0}=-10^{-1}$ and in both cases the rest of the parameters are as in the previous plots.

We also fix the spectral index $n_s$ and amplitude $A_s$ to $(n_s,A_s)=(1,2.3 \times 10^{-9})$, so that we can isolate the effects of the $f(R)$ models from the effects of a non-flat primordial spectrum. As we have mentioned in previous  sections, for large values of the parameter $b$ the HS model behaves as a matter dominated model and we actually expect the CMB spectrum at low multipoles to be nearly completely flat (also due to our choice of $n_s=1$).

In Fig.~\ref{fig:cmbcls} we present the low multipoles of the CMB TT power spectrum for the HS model (left panel) and the DES model (right panel). We compare several codes: our own modifications to CLASS which we call EFCLASS, the codes MGCAMB and FRCAMB for the HS model and the codes CLASS\_EOS\_FR and EFTCAMB for the DES model. We find that in the case of the DES model all approaches are in very good agreement, but in the case of the HS model, which also requires modifying the background evolution, there is significant disagreement at $l\in[2,5]$ as the codes MGCAMB and FRCAMB do not take into account the change of the background properly.

\begin{figure}[!t]
\centering
\includegraphics[width = 0.495\textwidth]{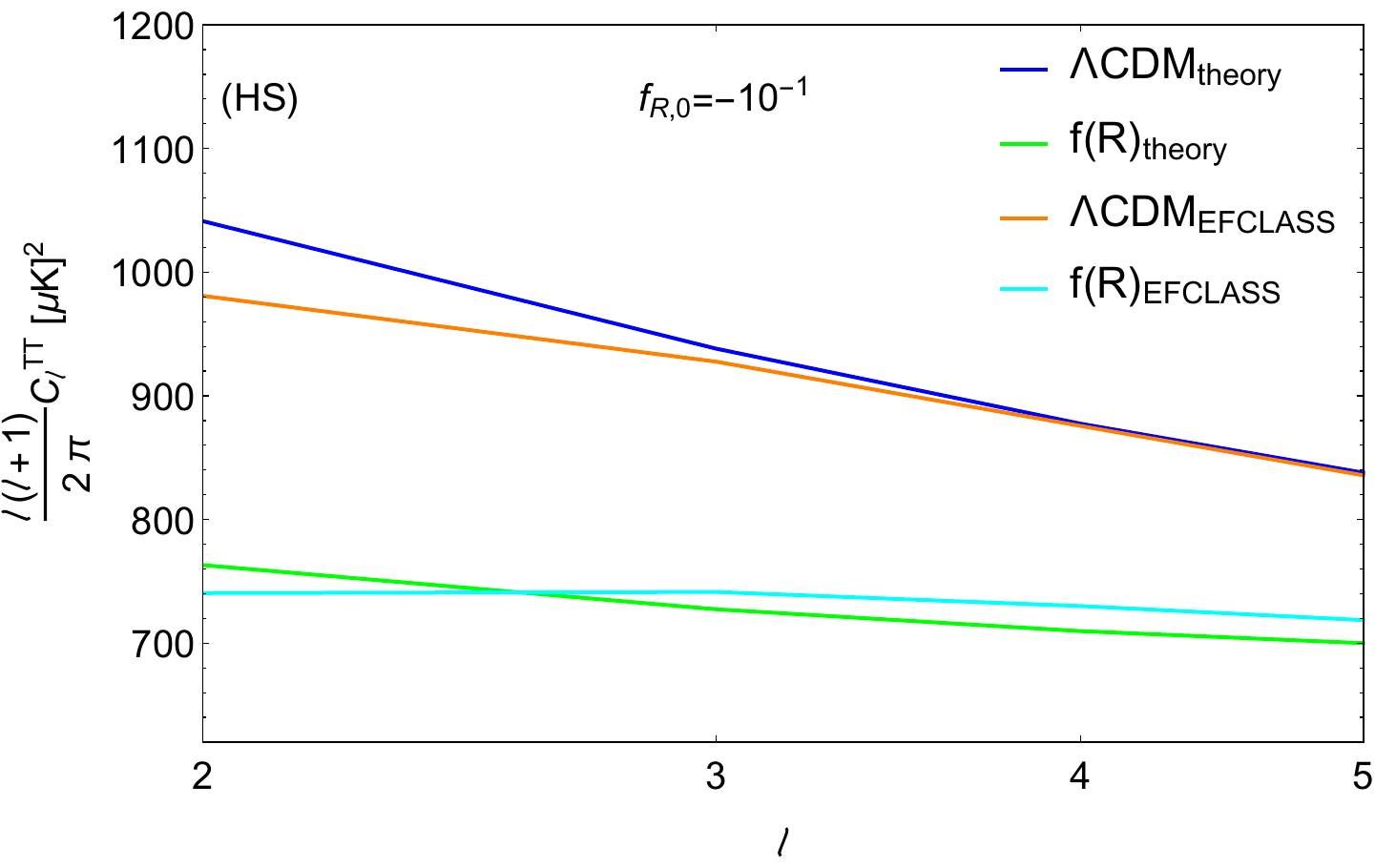}
\caption{A comparison of the low multipoles of the unlensed CMB TT power spectrum $(l\in[2,5])$ for the HS and \lcdm models between our own modifications to CLASS (EFCLASS) and a direct theoretical calculation using the expressions for the ISW effect given in Appendix \ref{Section:useful-formulae}. We find that in both cases there is excellent agreement. For this plot again we assume $(n_s,A_s)=(1,2.3 \times 10^{-9})$ and $f_{R,0}=-10^{-1}$, while the rest of the parameters are as in the previous plots.}
\label{fig:cmbclsISWtheory}
\end{figure}

Although disagreement between the codes for the HS model can be explained by the fact that the other codes do not treat the background properly, we also compare our results with a direct theoretical calculation of the ISW effect, see Fig.~\ref{fig:cmbclsISWtheory}. The relevant formulas for the theoretical calculation of the ISW effect are given for completeness in Appendix \ref{Section:useful-formulae}. In Fig.~\ref{fig:cmbclsISWtheory} we show the comparison of the low multipoles of the CMB TT power spectrum $(l\in[2,5])$ for the HS and \lcdm models between our own modifications to CLASS (EFCLASS) and a direct theoretical calculation. We find that in both cases there is excellent agreement. For this plot again we assume $(n_s,A_s)=(1,2.3 \times 10^{-9})$, $f_{R,0}=-10^{-1}$, while the rest of the parameters are as in the previous plots. We find that in the case of the HS model, the agreement between the direct theoretical calculation and our CLASS modifications (green and cyan lines respectively) is well below $\sim 2\%$.


\section{Evolution of the viscosity parameter}
\label{Section:viscosity}

In principle the anisotropic stress parameter is the lowest multipole in the Boltzmann hierarchy after the density and velocity perturbations. As a result, it should also follow an evolution equation. Since the properties of DE are currently unknown, one can assign a viscosity parameter $c_{vis}^2$ and a phenomenological evolution equation as in Ref. \cite{Hu:1998kj}:
\bea
\dot{\sigma}+3\mathcal{H}\frac{c_a^2}{w} \sigma&=&\frac83 \frac{c_{vis}^2}{1+w}\theta \nn \\
&=& \frac83 \frac{c_{vis}^2}{(1+w)^2}V_{DE},\label{eq:visc}
\eea
whereas in previous sections we have introduced the parameter $V_{DE}=(1+w)\theta$ and the adiabatic sound speed is $c_a^2=w-\frac{\dot{w}}{3\mathcal{H}(1+w)}=w-\frac{a w'}{3(1+w)}$ where dots are conformal time derivatives and primes scale factor derivatives. Also, note that there is a difference in the definition of the anisotropic stress compared to Ref.~\cite{Hu:1998kj}. Since we follow the notation of Ref.~\cite{Ma:1995ey} we have $\pi_{DE}=w\Pi_{\textrm{WH}}$, where $\pi_{DE}=\frac32(1+w)\sigma$ is the anisotropic stress in this paper and $\Pi_{\textrm{WH}}$ is the anisotropic stress parameter of Ref.~\cite{Hu:1998kj}.

The parameterization of Eq.~(\ref{eq:visc}) is also useful if one wants to explore the properties of a generalized dark matter fluid, as was done in Ref.~\cite{Kunz:2016yqy} or place constraints in imperfect fluids \cite{Mota:2007sz}. In our case we actually know the underlying DE model, which is our $f(R)$ effective fluid, so using Eq.~(\ref{eq:visc}) we can reconstruct the viscosity parameter, something which would be of great interest for forecasts for upcoming surveys.

After changing variables from conformal time to scale factor in Eq.~(\ref{eq:visc}) we can solve for the viscosity parameter as:
\be
c_{vis}^2=\frac{a H (1+w)}{4 V_{DE} w}\left(3 c_a^2(1+w)\pi_{DE}+w(a \pi_{DE}'-3 w \pi_{DE})\right).\label{eq:cviseq}
\ee

In the case of the HS model it can easily be seen from the previous equation that at early times, in matter domination in particular, the viscosity parameter scales as
\be
c_{vis}^2 \simeq \frac{14}{3} \frac{1-\Omega_{m0}}{\Omega_{m0}^2}\;b\;k^2\;a^4.
\ee
In the case of the DES model, we have that while $c_{vis}^2\rightarrow0$ there is clearly anisotropic stress in this model as in the RHS of Eq.~(\ref{eq:visc}) the term $(1+w)$ in the denominator cancels out with $c_{vis}^2$ to give a non-zero result.

\begin{figure}[!t]
\centering
\includegraphics[width = 0.495\textwidth]{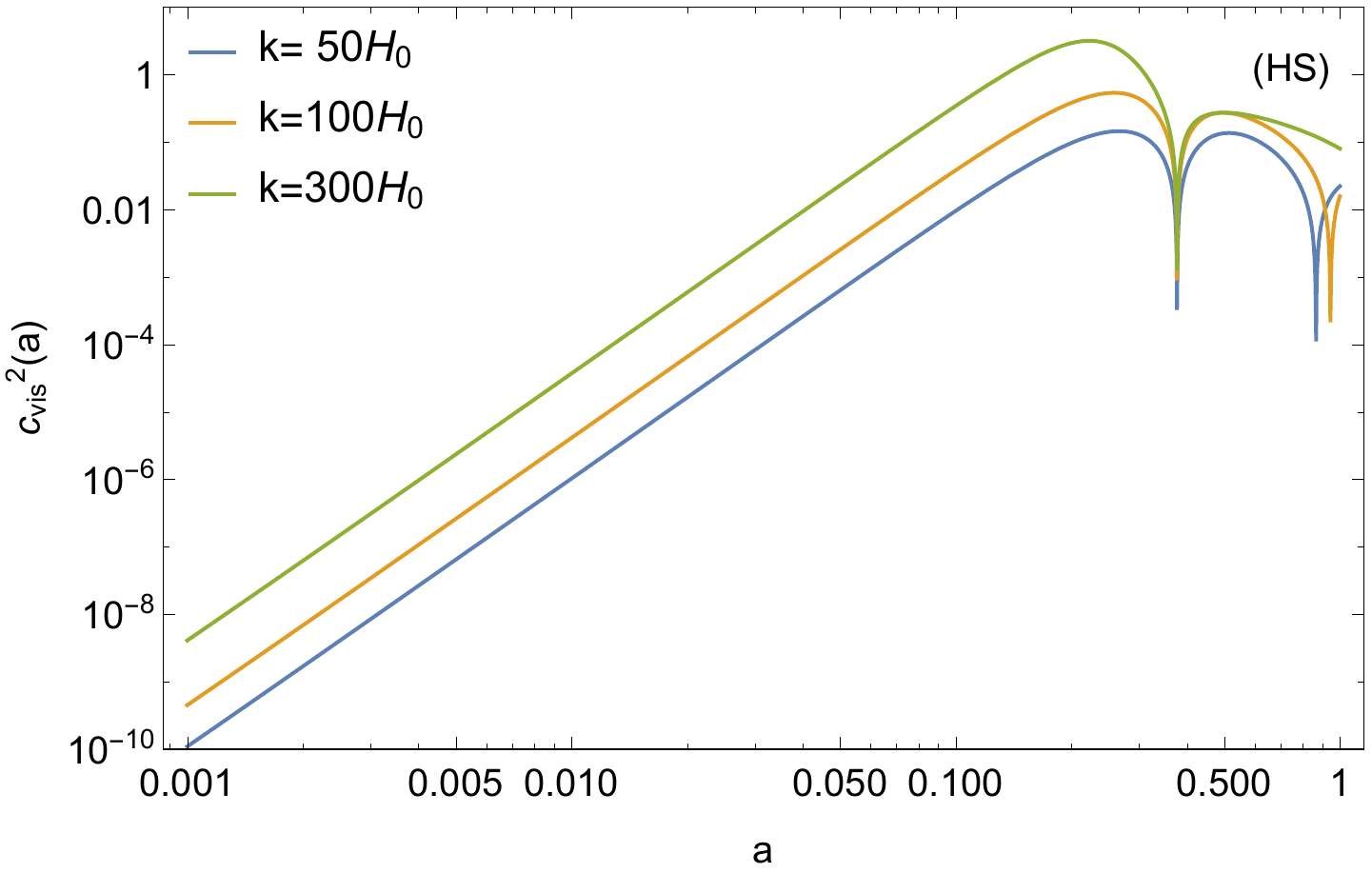}
\caption{The evolution of the viscosity $c_{vis}^2(a)$ parameter for the HS model for $\Omega_{m0}=0.3$, $f_{R,0}=-10^{-4}$ and values of the wavenumber $k/H_0=[50,100,300]$. As can be seen, the parameter changes by more than 7 orders of magnitude over the range $a\in[10^{-3},1]$.}
\label{fig:viscocity}
\end{figure}

In Fig.~\ref{fig:viscocity} we show the evolution of the viscosity parameter $c_{vis}^2$ given by Eq.~(\ref{eq:cviseq}) as a function of scale factor $a$ for the HS model for $\Omega_{m0}=0.3$, $f_{R,0}=-10^{-4}$ and values of the wave number $k/H_0=[50,100,300]$. As can be seen, the parameter changes by more than 7 orders of magnitude over the range $a\in[10^{-3},1]$ which means that in realistic models, like the HS $f(R)$ model, $c_{vis}^2$ clearly cannot be considered as a constant parameter, as is the usual assumption when performing forecasts for future surveys like Euclid \cite{Sapone:2013wda}. 

\section{Cosmological constraints}
\label{Section:mcmc-results}

\subsection{Data}

Here we present the results of our analysis from fitting the latest cosmological observations including the supernovae type Ia (SnIa), Baryon Acoustic Oscillations (BAO), CMB, the Hubble expansion H(z) and growth $f\sigma_8$ data. In particular, we use the Pantheon SnIa data of Ref.~\cite{Scolnic:2017caz}, the BAO points from 6dFGS \cite{Beutler:2011hx}, SDDS \cite{Anderson:2013zyy}, BOSS CMASS \cite{Xu:2012hg}, WiggleZ \cite{Blake:2012pj}, MGS \cite{Ross:2014qpa} and BOSS DR12 \cite{Gil-Marin:2015nqa}. We also use the CMB shift parameters based on the \textit{Planck 2015} release \cite{Ade:2015xua}, as derived by Ref.~\cite{Wang:2015tua}.\footnote{As of writing, the likelihoods of the Planck 2018 data release are not publicly available.}

The Hubble expansion $H(z)$ data are in general derived in two ways: by the differential age method and by the clustering of galaxies or quasars. The former is related to the redshift drift of distant objects over significant time periods, usually a decade or longer, since in GR the Hubble parameter can also be written in terms of the rate of change of the redshift $H(z)=-\frac{1}{1+z}\frac{dz}{dt}$ \cite{Jimenez:2001gg}. The latter approach is related to the clustering of galaxies or quasars and it provides direct measurements of the Hubble parameter by measuring the BAO peak in the radial direction \cite{Gaztanaga:2008xz}. The compilation of Hubble parameter $H(z)$ data that we used in our analysis are shown in Table \ref{tab:Hzdata} along with the corresponding references.

We use the growth-rate data compilation of Ref.~\cite{Sagredo:2018ahx} which is presented in Table~\ref{tab:fs8tab} with the corresponding references. In Ref.~\cite{Sagredo:2018ahx} the authors analyzed combinations of subsets in the dataset and used Bayesian model comparison to show that this particular dataset is internally robust. The growth-rate data used in our analysis come from measurements of redshift-space distortions, which are probes of the Large Scale Structure (LSS) and in fact measure the combination $f\sigma_8(a)\equiv f(a)\cdot \sigma(a)$, where $f(a)=\frac{d ln\delta}{d lna}$ is the growth rate, $\sigma(a)=\sigma_{8,0}\frac{\delta(a)}{\delta(1)}$  is  the redshift-dependent rms fluctuations of the linear density field within spheres of radius $R=8 h^{-1} \textrm{\textrm{Mpc}}$, and the parameter $\sigma_{8,0}$ is its value today.

It is important to stress that $f\sigma_8(a)$ can be measured directly from the ratio of the monopole to the quadrupole of the redshift-space power spectrum. This depends on the combination $\beta=f/b_0$, where $f$ is the growth rate and $b_0$ is the bias, with all quantities  defined assuming linear theory \cite{Percival:2008sh,Song:2008qt,Nesseris:2006er}. Then, $f\sigma_8(a)$ can be shown to be independent of bias, as in this combination it completely cancels out. Indeed, this combination has been shown to be a good discriminator of DE models \cite{Song:2008qt}. For details on the covariances of the data and how to properly correct for the Alcock-Paczynski effect see Refs.~\cite{Sagredo:2018ahx}, \cite{Nesseris:2017vor} and \cite{Kazantzidis:2018rnb}, while for previous related analyses see Refs.~\cite{Basilakos:2018arq,Basilakos:2017rgc,Basilakos:2016nyg}.

\subsection{Methodology}

Our total likelihood function $L_{\rm tot}$ can be given as the product of the various likelihoods as

$$
L_{\rm tot}=L_{\rm SnIa} \times L_{\rm BAO} \times L_{\rm H(z)} \times L_{\rm cmb}
\times L_{\rm growth},
$$
which can also be translated to the total $\chi^2$ via $\chi^{2}_{\rm tot}=-2\log{L_{\rm tot}}$ or
\be
\chi^{2}_{\rm tot}=\chi^{2}_{\rm SnIa}+\chi^{2}_{\rm BAO}+\chi^{2}_{\rm H(z)}+
\chi^{2}_{\rm cmb}+\chi^{2}_{\rm growth} .\label{eq:chi2eq}
\ee

In order to study the statistical significance of our constraints we will use the well known Akaike Information Criterion (AIC)~\cite{Akaike1974}. Assuming Gaussian errors the AIC estimator is given by
\begin{eqnarray}
{\rm AIC} = -2 \ln {\cal L}_{\rm max}+2k_p+\frac{2k_p(k_p+1)}{N_{\rm dat}-k_p-1} \label{eq:AIC}\;,
\end{eqnarray}
where $N_{\rm dat}$ and $k_p$ indicate the total number of data points and the number of free parameters (see also~\cite{Liddle:2007fy}) of our models, respectively. In our case we have 1048 data points from the Pantheon set, 3 CMB shift parameters, 9 BAO points, 22 growth-rate data and 36 $H(z)$ points for a total of $N_{\rm dat}=1118$.

The usual interpretation of the AIC estimator is that a smaller value implies a better fit to the data. However, in order to compare different models, we need to use the pair difference which can be written as $\Delta {\rm  AIC}={\rm AIC}_{\rm model}-{\rm AIC}_{\rm min}$. This relative difference can be interpreted with the Jeffreys' scale as follows: $4<\Delta {\rm AIC} <7$ indicate a positive evidence against the model with higher value of ${\rm AIC}_{\rm model}$ and $\Delta {\rm AIC} \ge 10$ suggests strong evidence. Finally, when we have that $\Delta {\rm AIC} \le 2$ then this is interpreted as an indication of the consistency of the two models. However, note that the Jeffreys' scale in general has been shown to lead to misleading conclusions, thus it has to be interpreted with care \cite{Nesseris:2012cq}.

To summarize, our $\chi^2$ is given by Eq.~(\ref{eq:chi2eq}) and the parameter vectors (assuming a flat Universe) are given by: $p_{\Lambda \textrm{CDM}}=\left(\Omega_{m0}, 100\Omega_b h^2, h, \sigma_{8,0}\right)$ for the \lcdm; and  $p_{f(R)}=\left(\Omega_{m0}, 100\Omega_b h^2, \alpha, h, \sigma_{8,0}\right)$ for the $f(R)$ models (when studying the DES model $\alpha = f_{R,0}$ whereas for the HS model $\alpha=b$). Then, the best-fit parameters and their uncertainties were obtained via the MCMC method based on a Metropolis-Hastings algorithm written by one of the authors.\footnote{The MCMC code for Mathematica used in the analysis is freely available at \url{http://members.ift.uam-csic.es/savvas.nesseris/}.} Moreover, we assumed priors for the parameters given by $\Omega_{m0} \in[0.1, 0.5]$, $\Omega_b h^2 \in[0.001, 0.08]$, $\alpha=(-f_{R,0},b) \in[0, 1]$, $h \in[0.4, 1]$, $\sigma_{8,0}\in[0.1, 1.8]$ and obtained approximately $\sim10^5$ points for each of the three models.

\begin{table}[!t]
\caption{The $H(z)$ data used in the current analysis (in units of $\textrm{km}~\textrm{s}^{-1} \textrm{Mpc}^{-1}$). This compilation is partly based on those of Refs.~\cite{Moresco:2016mzx} and \cite{Guo:2015gpa}.\label{tab:Hzdata}}
\small
\centering
\begin{tabular}{cccccccccc}
\\
\hline\hline
$z$  & $H(z)$ & $\sigma_{H}$ & Ref.   \\
\hline
$0.07$    & $69.0$   & $19.6$  & \cite{Zhang:2012mp}  \\
$0.09$    & $69.0$   & $12.0$  & \cite{STERN:2009EP} \\
$0.12$    & $68.6$   & $26.2$  & \cite{Zhang:2012mp}  \\
$0.17$    & $83.0$   & $8.0$   & \cite{STERN:2009EP}    \\
$0.179$   & $75.0$   & $4.0$   & \cite{MORESCO:2012JH}   \\
$0.199$   & $75.0$   & $5.0$   & \cite{MORESCO:2012JH}   \\
$0.2$     & $72.9$   & $29.6$  & \cite{Zhang:2012mp}   \\
$0.27$    & $77.0$   & $14.0$  & \cite{STERN:2009EP}   \\
$0.28$    & $88.8$   & $36.6$  & \cite{Zhang:2012mp}  \\
$0.35$    & $82.7$   & $8.4$   & \cite{Chuang:2012qt}   \\
$0.352$   & $83.0$   & $14.0$  & \cite{MORESCO:2012JH}   \\
$0.3802$  & $83.0$   & $13.5$  & \cite{Moresco:2016mzx}   \\
$0.4$     & $95.0$   & $17.0$  & \cite{STERN:2009EP}    \\
$0.4004$  & $77.0$   & $10.2$  & \cite{Moresco:2016mzx}   \\
$0.4247$  & $87.1$   & $11.2$  & \cite{Moresco:2016mzx}   \\
$0.44$    & $82.6$   & $7.8$   & \cite{Blake:2012pj}   \\
$0.44497$ & $92.8$   & $12.9$  & \cite{Moresco:2016mzx}   \\
$0.4783$  & $80.9$   & $9.0$   & \cite{Moresco:2016mzx}   \\
\hline\hline
\end{tabular}
\begin{tabular}{cccccccccc}
\\
\hline\hline
$z$  & $H(z)$ & $\sigma_{H}$ & Ref.   \\
\hline
$0.48$    & $97.0$   & $62.0$  & \cite{STERN:2009EP}   \\
$0.57$    & $96.8$   & $3.4$   & \cite{Anderson:2013zyy}   \\
$0.593$   & $104.0$  & $13.0$  & \cite{MORESCO:2012JH}  \\
$0.60$    & $87.9$   & $6.1$   & \cite{Blake:2012pj}   \\
$0.68$    & $92.0$   & $8.0$   & \cite{MORESCO:2012JH}    \\
$0.73$    & $97.3$   & $7.0$   & \cite{Blake:2012pj}   \\
$0.781$   & $105.0$  & $12.0$  & \cite{MORESCO:2012JH} \\
$0.875$   & $125.0$  & $17.0$  & \cite{MORESCO:2012JH} \\
$0.88$    & $90.0$   & $40.0$  & \cite{STERN:2009EP}   \\
$0.9$     & $117.0$  & $23.0$  & \cite{STERN:2009EP}   \\
$1.037$   & $154.0$  & $20.0$  & \cite{MORESCO:2012JH} \\
$1.3$     & $168.0$  & $17.0$  & \cite{STERN:2009EP}   \\
$1.363$   & $160.0$  & $33.6$  & \cite{Moresco:2015cya}  \\
$1.43$    & $177.0$  & $18.0$  & \cite{STERN:2009EP}   \\
$1.53$    & $140.0$  & $14.0$  & \cite{STERN:2009EP}  \\
$1.75$    & $202.0$  & $40.0$  & \cite{STERN:2009EP}  \\
$1.965$   & $186.5$  & $50.4$  & \cite{Moresco:2015cya}  \\
$2.34$    & $222.0$  & $7.0$   & \cite{Delubac:2014aqe}   \\
\hline\hline
\end{tabular}
\end{table}

\begin{table}[!t]
\caption[]{Compilation of the $f\sigma_8(z)$ measurements used in this analysis along with the reference matter density parameter $\Omega_{m_0}$ (needed for the growth correction) and related references.	 \label{tab:fs8tab}}
\begin{center}
\begin{tabular}{ccccccccc}
\hline
\hline
$z$     & $f\sigma_8(z)$ & $\sigma_{f\sigma_8}(z)$  & $\Omega_{m,0}^\text{ref}$ & Ref. \\ \hline
0.02    & 0.428 & 0.0465  & 0.3 & \cite{Huterer:2016uyq}   \\
0.02    & 0.398 & 0.065   & 0.3 & \cite{Turnbull:2011ty},\cite{Hudson:2012gt} \\
0.02    & 0.314 & 0.048   & 0.266 & \cite{Davis:2010sw},\cite{Hudson:2012gt}  \\
0.10    & 0.370 & 0.130   & 0.3 & \cite{Feix:2015dla}  \\
0.15    & 0.490 & 0.145   & 0.31 & \cite{Howlett:2014opa}  \\
0.17    & 0.510 & 0.060   & 0.3 & \cite{Song:2008qt}  \\
0.18    & 0.360 & 0.090   & 0.27 & \cite{Blake:2013nif} \\
0.38    & 0.440 & 0.060   & 0.27 & \cite{Blake:2013nif} \\
0.25    & 0.3512 & 0.0583 & 0.25 & \cite{Samushia:2011cs} \\
0.37    & 0.4602 & 0.0378 & 0.25 & \cite{Samushia:2011cs} \\
0.32    & 0.384 & 0.095  & 0.274 & \cite{Sanchez:2013tga}   \\
0.59    & 0.488  & 0.060 & 0.307115 & \cite{Chuang:2013wga} \\
0.44    & 0.413  & 0.080 & 0.27 & \cite{Blake:2012pj} \\
0.60    & 0.390  & 0.063 & 0.27 & \cite{Blake:2012pj} \\
0.73    & 0.437  & 0.072 & 0.27 & \cite{Blake:2012pj} \\
0.60    & 0.550  & 0.120 & 0.3 & \cite{Pezzotta:2016gbo} \\
0.86    & 0.400  & 0.110 & 0.3 & \cite{Pezzotta:2016gbo} \\
1.40    & 0.482  & 0.116 & 0.27 & \cite{Okumura:2015lvp} \\
0.978   & 0.379  & 0.176 & 0.31 & \cite{Zhao:2018jxv} \\
1.23    & 0.385  & 0.099 & 0.31 & \cite{Zhao:2018jxv} \\
1.526   & 0.342  & 0.070 & 0.31 & \cite{Zhao:2018jxv} \\
1.944   & 0.364  & 0.106 & 0.31 & \cite{Zhao:2018jxv} \\
\hline
\hline
\end{tabular}
\end{center}
\end{table}

\begin{table}[!t]
\begin{center}
\caption{$\Lambda$CDM parameters with $68\%$ limits based on TT,TE,EE+lowP and a flat $\Lambda$CDM model (middle column) or a $w$CDM model (right column); see Table 4 of Ref.~\cite{Ade:2015xua} and the Planck chains archive.\label{tab:planck}}
\begin{tabular}{ccc}\hline \hline
Parameter & Value ($\Lambda$CDM) & Value ($w$CDM) \\
\hline
$\Omega_b h^2$ & $0.02225\pm0.00016$ & $0.02229\pm0.00016$ \\
$\Omega_c h^2$ & $0.1198\pm0.0015$ & $0.1196\pm0.0015$\\
$n_s$ & $0.9645\pm0.0049$ & $0.9649\pm0.0048$\\
$H_0$ & $67.27\pm0.66$ & $>81.3$\\
$\Omega_m$ & $0.3156\pm0.0091$ & $0.203^{+0.022}_{-0.065}$\\
$w$ & $-1$ & $-1.55^{+0.19}_{-0.38}$\\
$\sigma_8$ & $0.831\pm0.013$ & $0.983^{+0.100}_{-0.055}$\\
\hline \hline
\end{tabular}
\end{center}
\end{table}

\subsection{Results}

\begin{figure*}[!t]
\centering
\includegraphics[width = 0.48\textwidth]{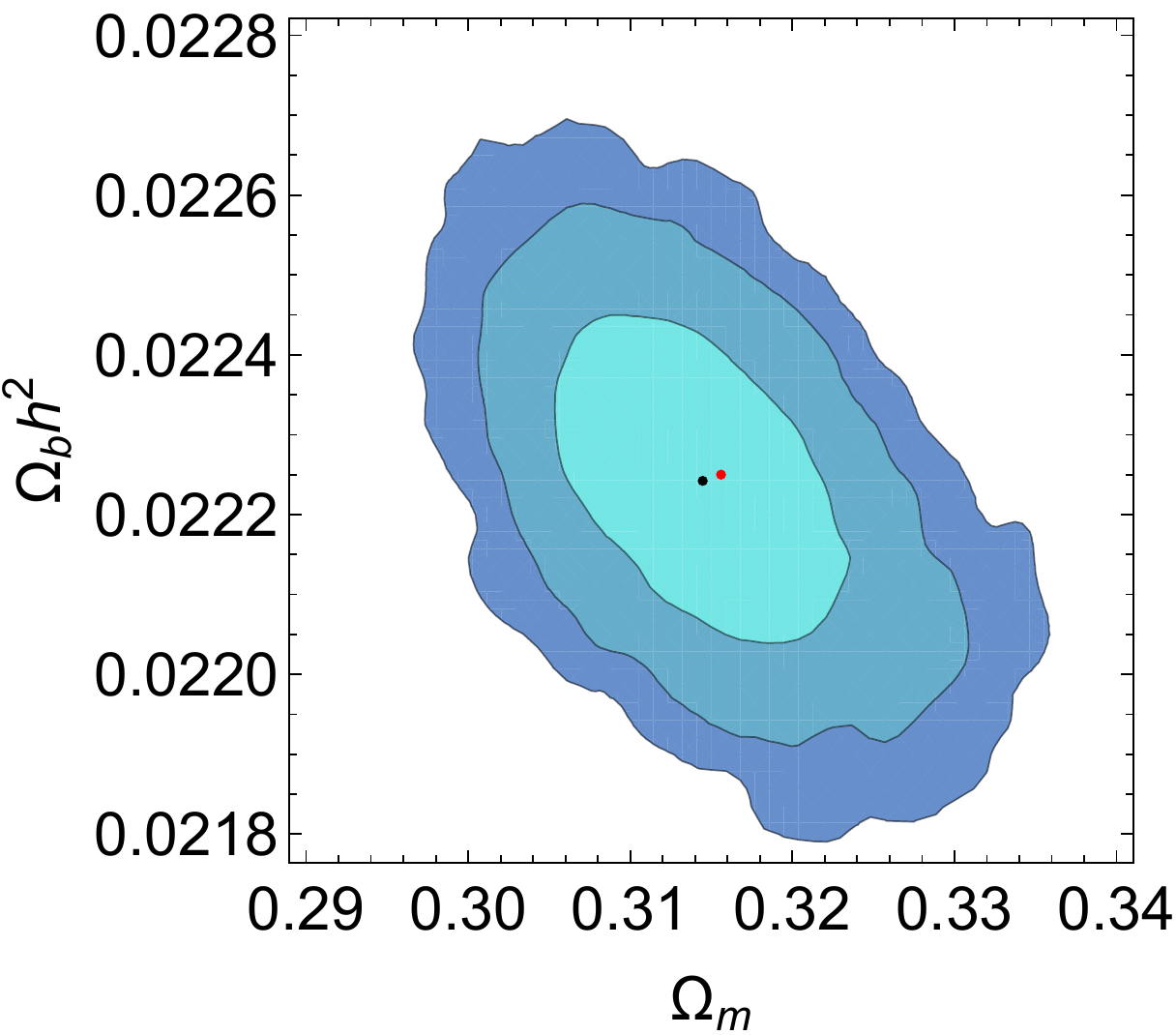}
\includegraphics[width = 0.46\textwidth]{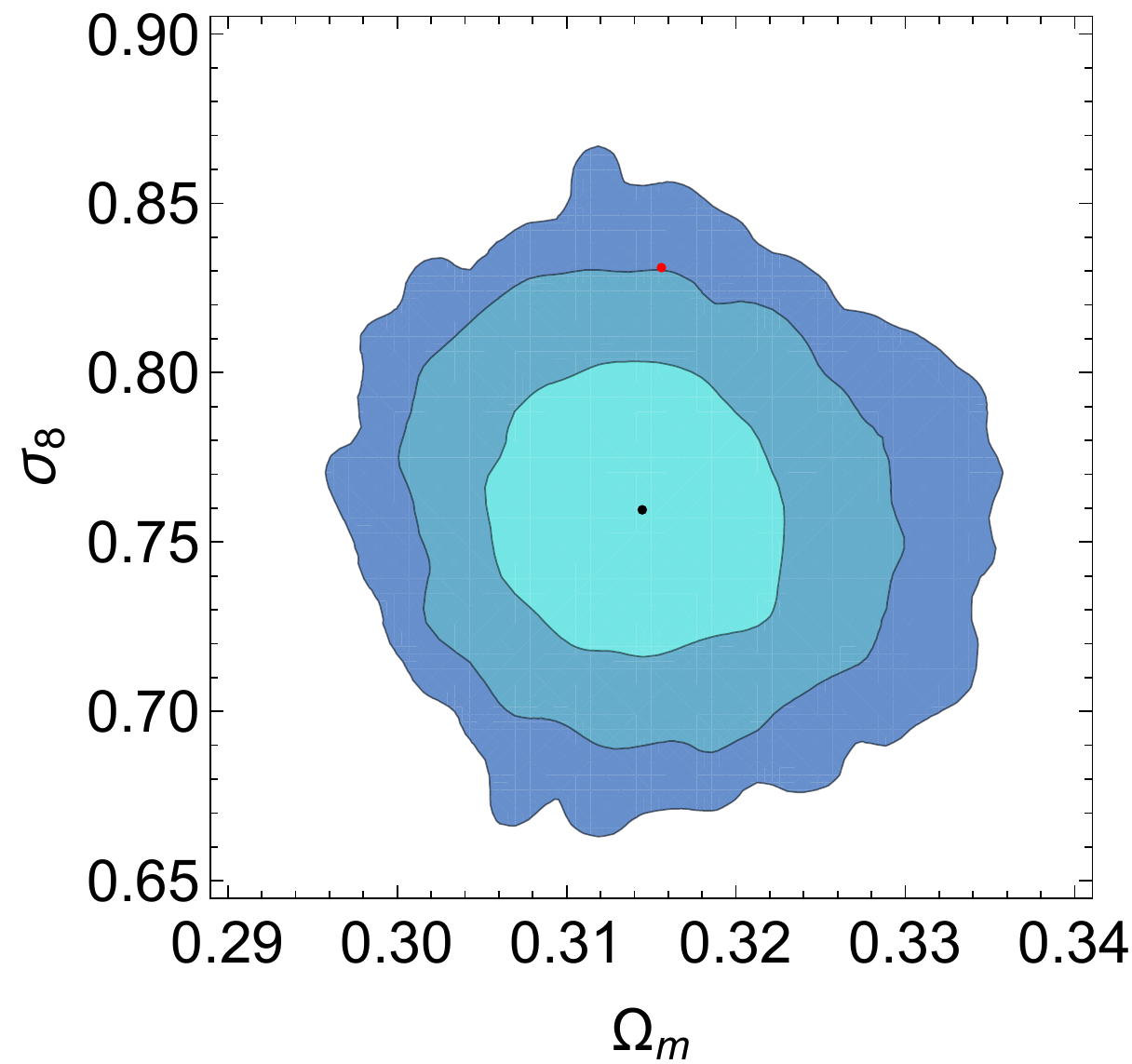}
\includegraphics[width = 0.85\textwidth]{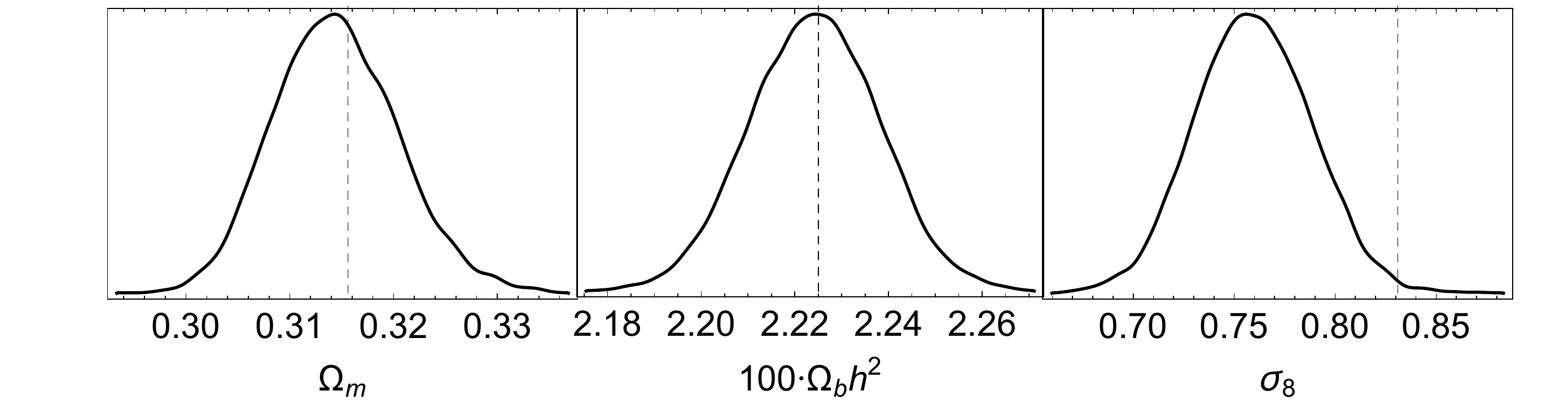}
\caption{The 68.3$\%$, 95.4$\%$ and 99.7$\%$ confidence contours (top) and the one-dimensional marginalized likelihoods (bottom) for various parameter combinations for the $\Lambda$CDM model. The red point and black dashed lines correspond to the concordance Planck 2015 $\Lambda$CDM parameters given in Table \ref{tab:planck}. The black point indicates the mean value from the MCMC analysis. \label{fig:MCMCLCDM}}
\end{figure*}

\begin{figure*}[!t]
\centering
\includegraphics[width = 0.34\textwidth]{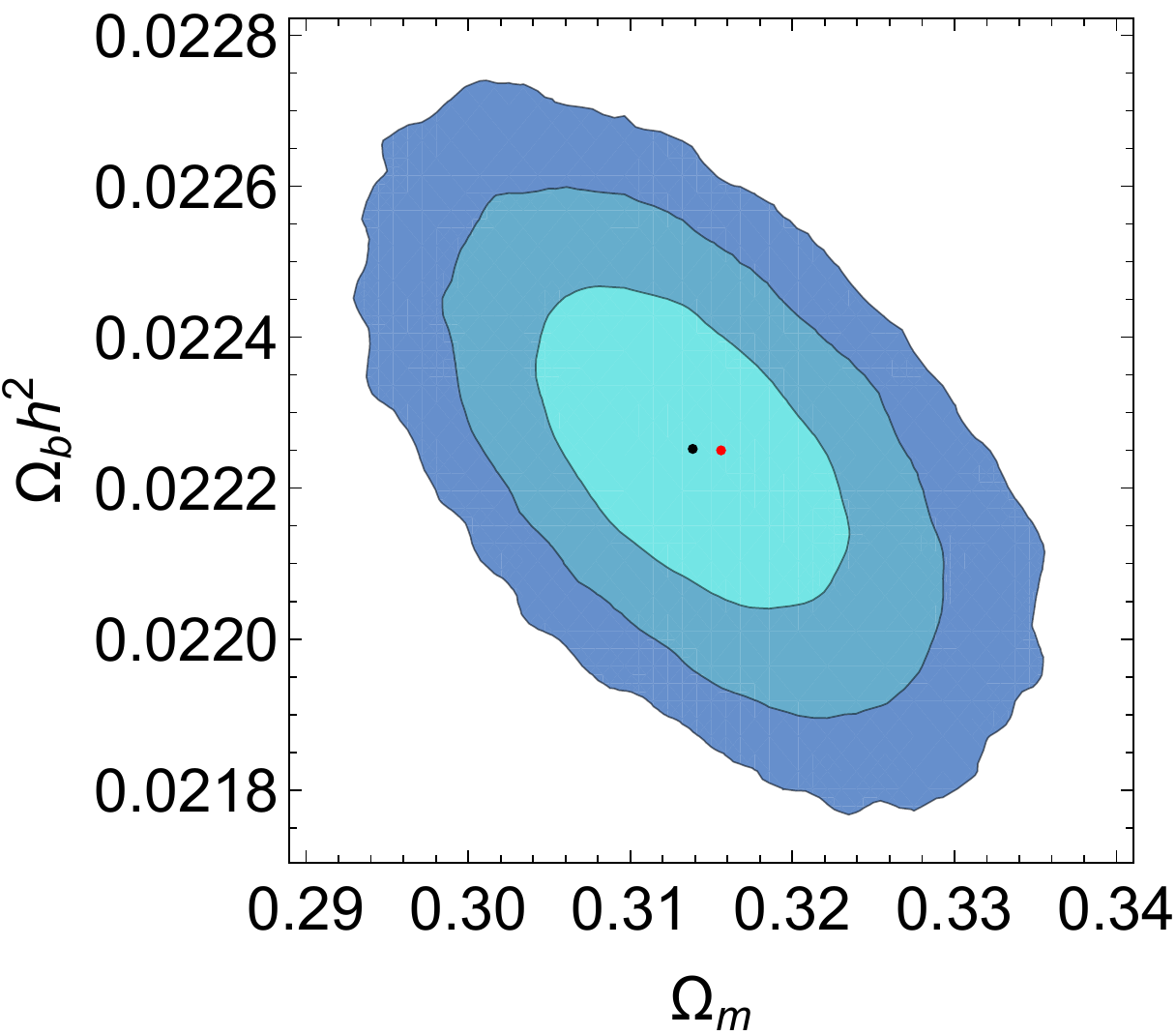}
\includegraphics[width = 0.31\textwidth]{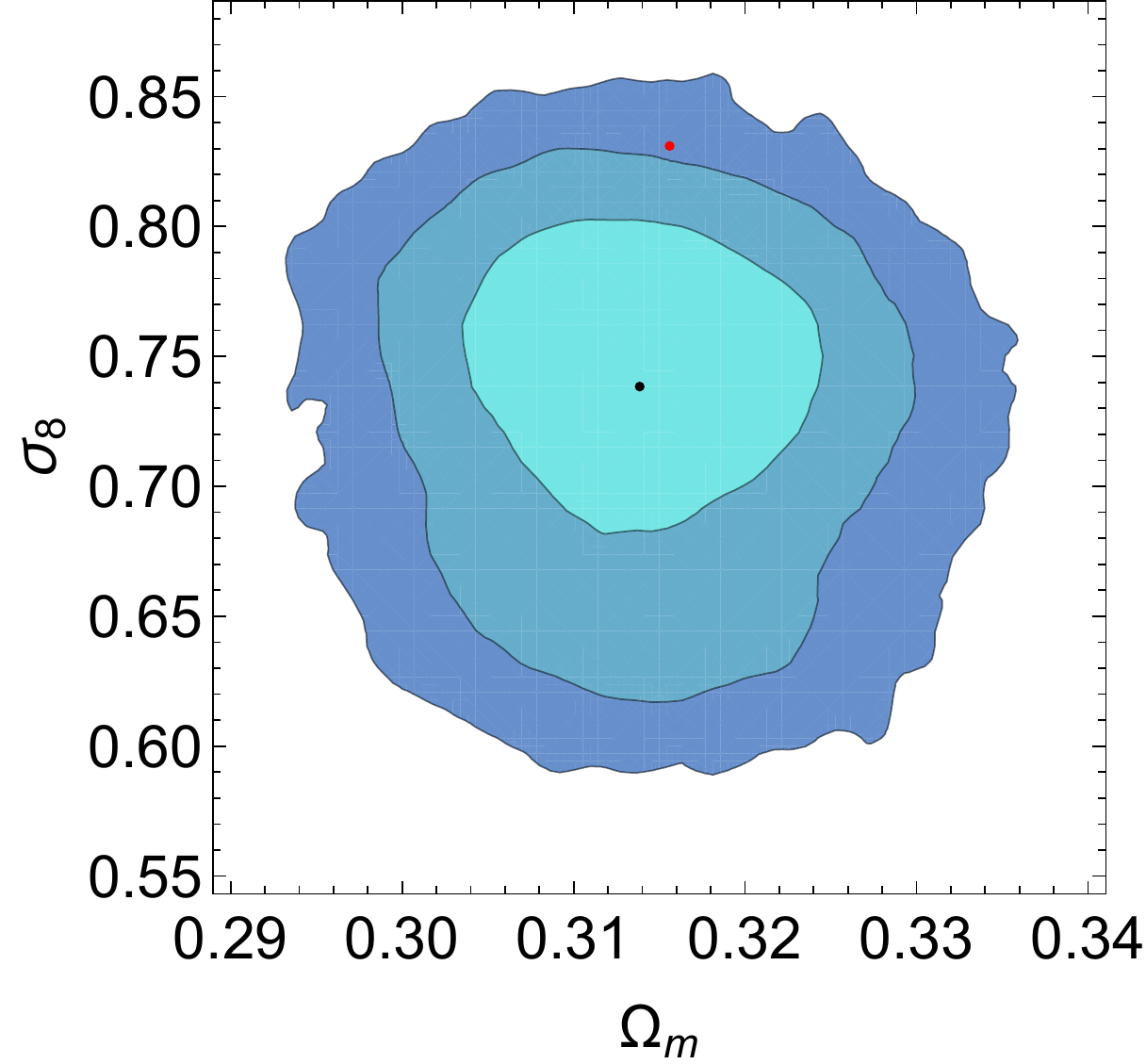}
\includegraphics[width = 0.30\textwidth]{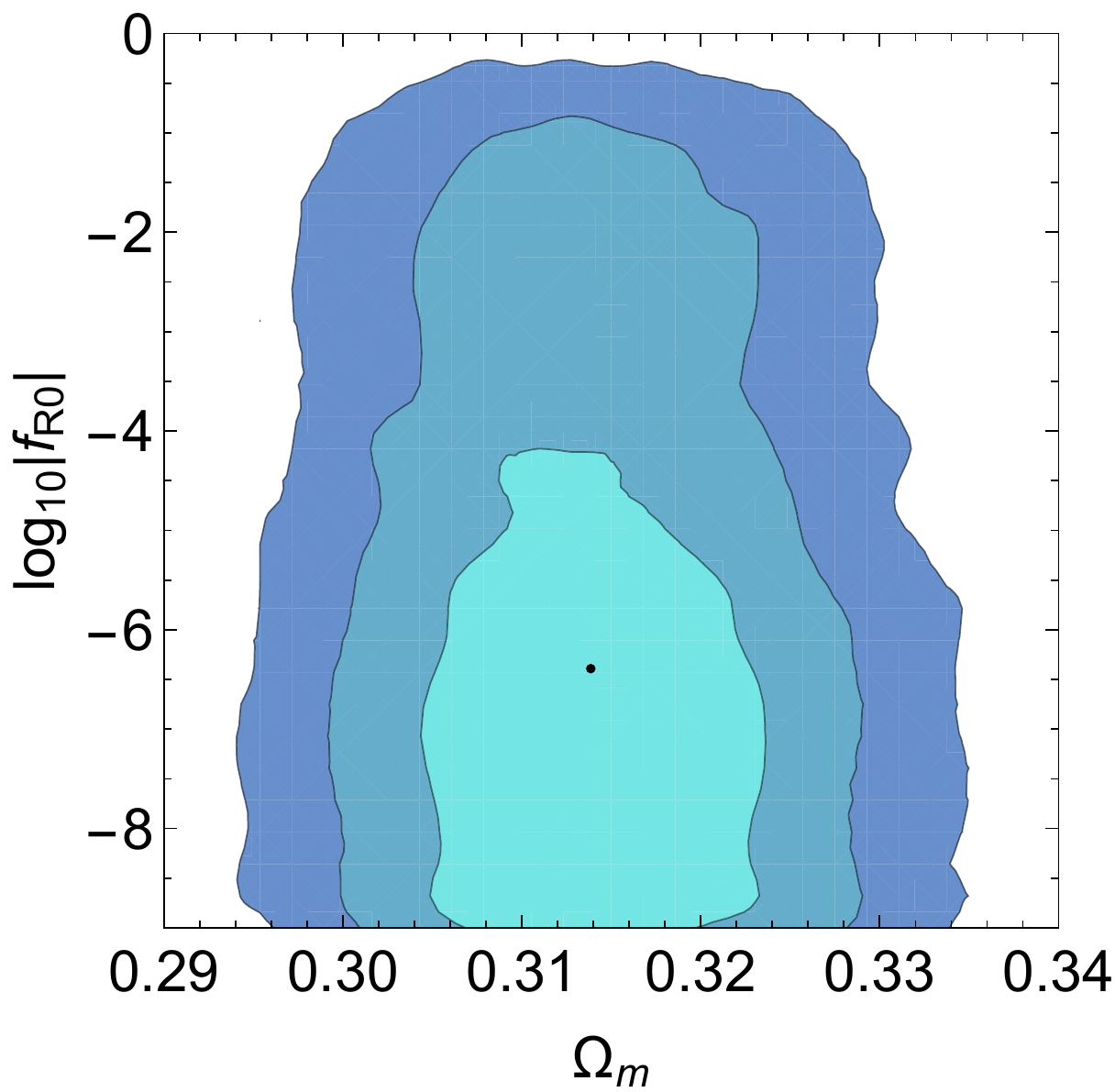}
\includegraphics[width = 0.85\textwidth]{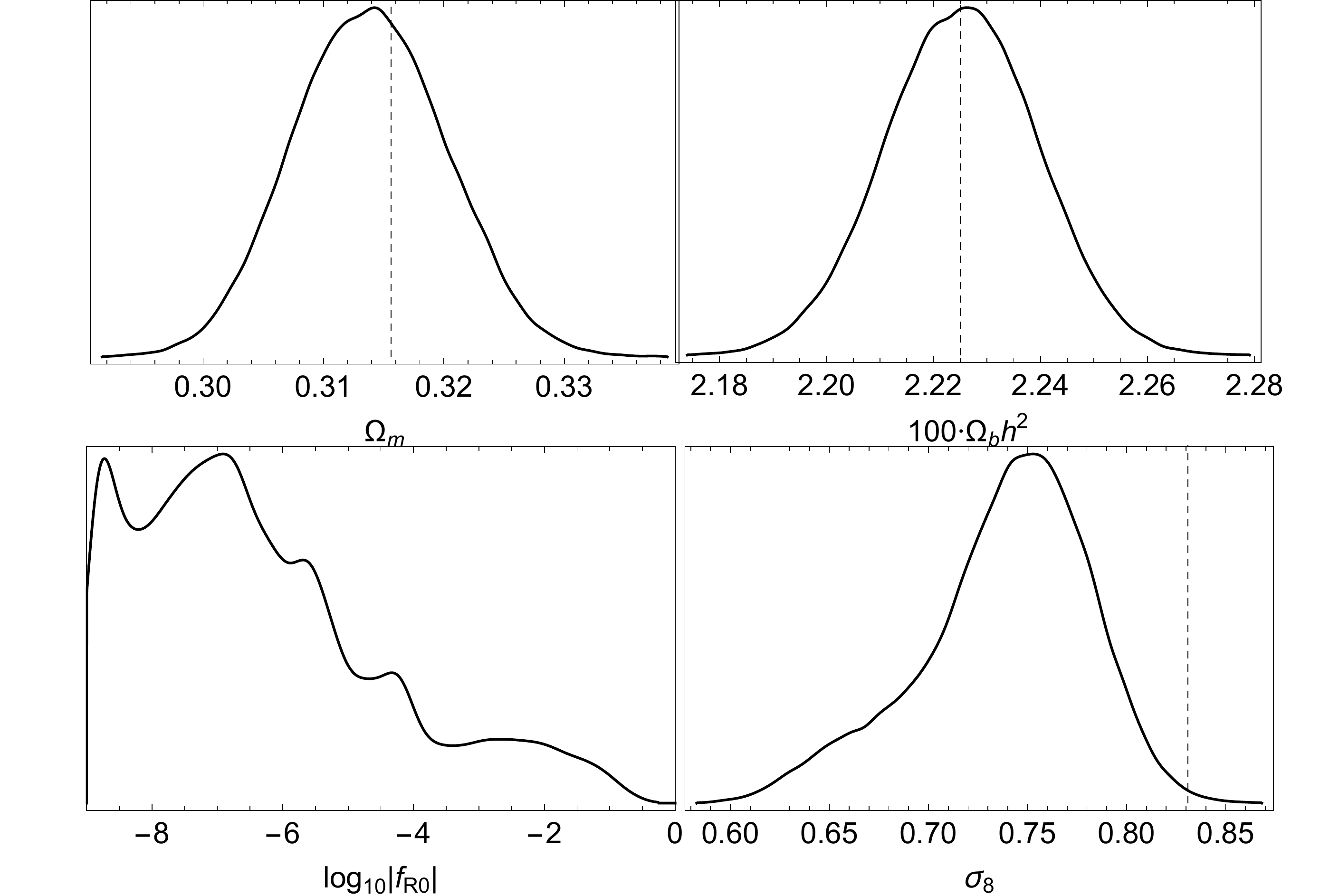}
\caption{The 68.3$\%$, 95.4$\%$ and 99.7$\%$ confidence contours (top) and the one-dimensional marginalized likelihoods (bottom) for various parameter combinations for the DES model. The red point and black dashed lines correspond to the concordance Planck 2015 $\Lambda$CDM parameters given in Table \ref{tab:planck}. The black point indicates the mean value from the MCMC analysis. \label{fig:MCMCDES}}
\end{figure*}

\begin{figure*}[!t]
\centering
\includegraphics[width = 0.34\textwidth]{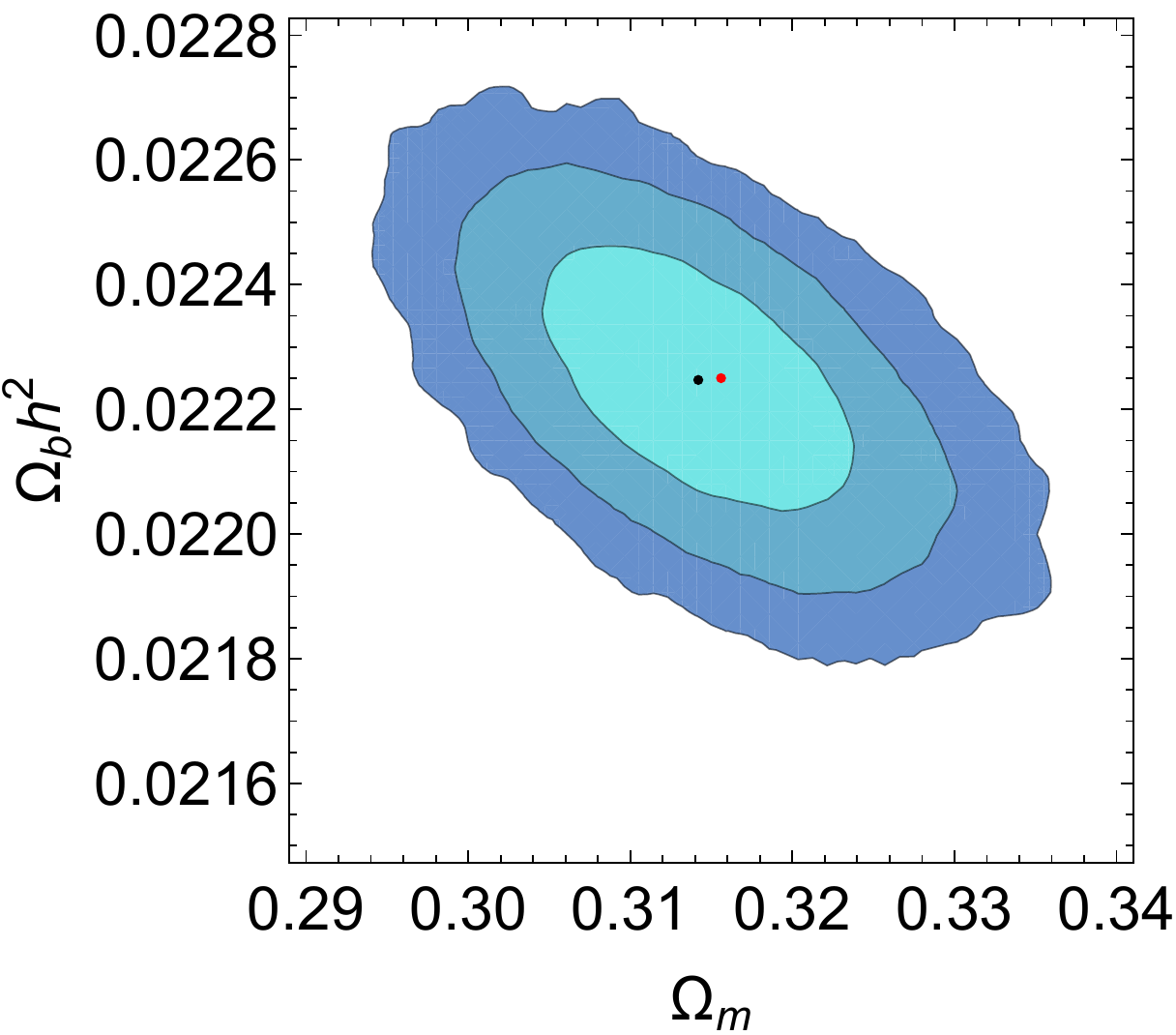}
\includegraphics[width = 0.31\textwidth]{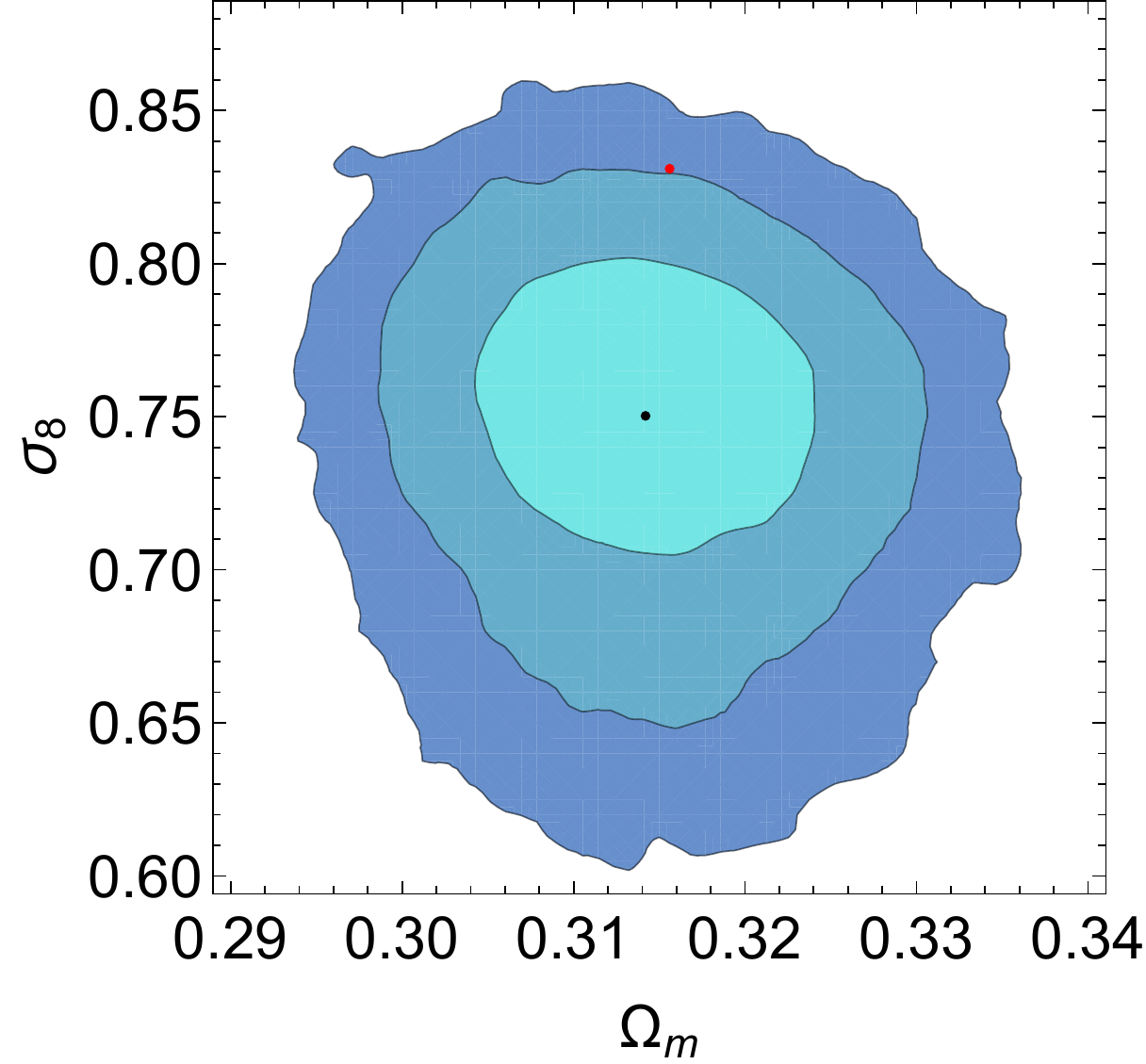}
\includegraphics[width = 0.30\textwidth]{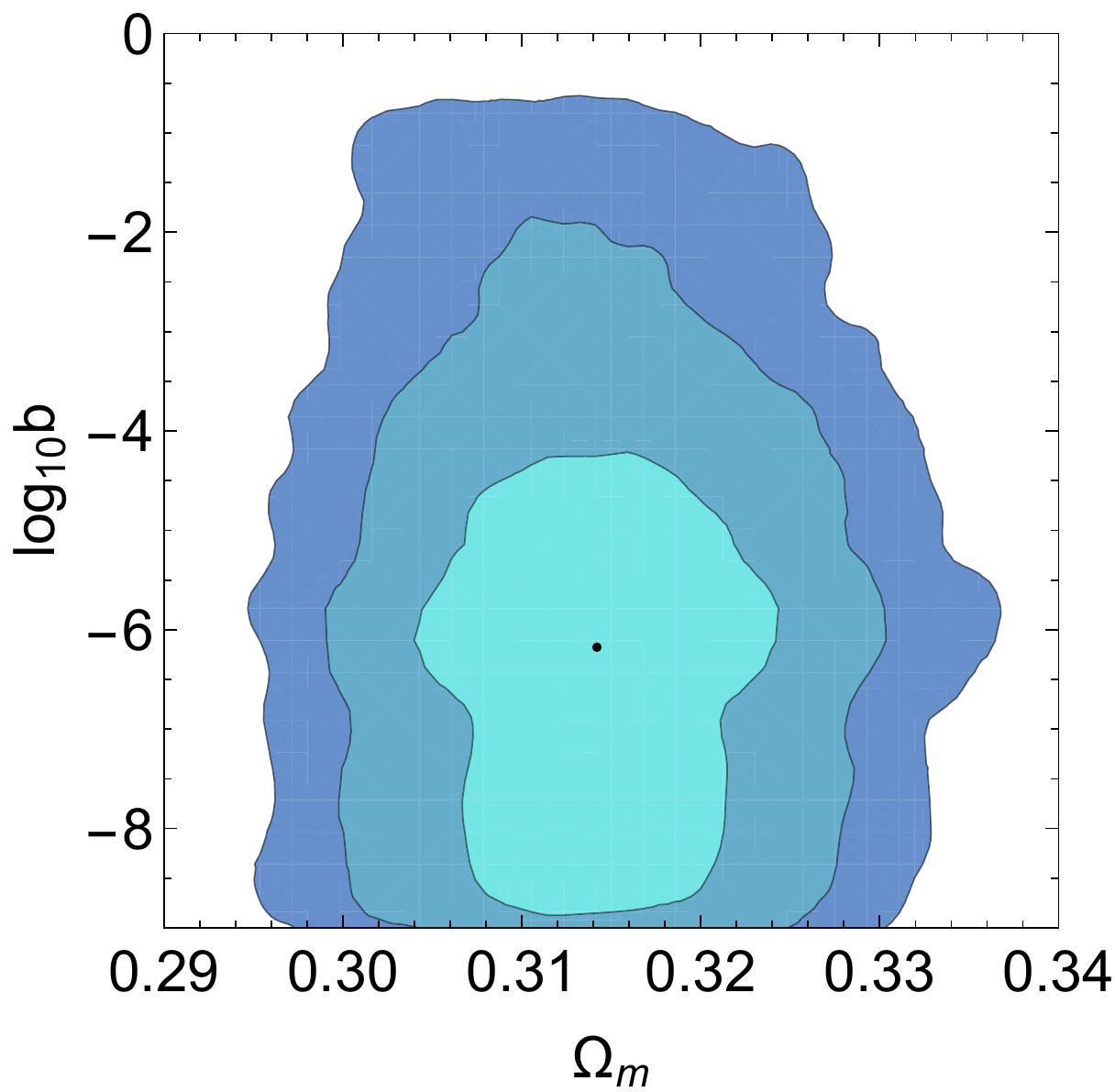}
\includegraphics[width = 0.85\textwidth]{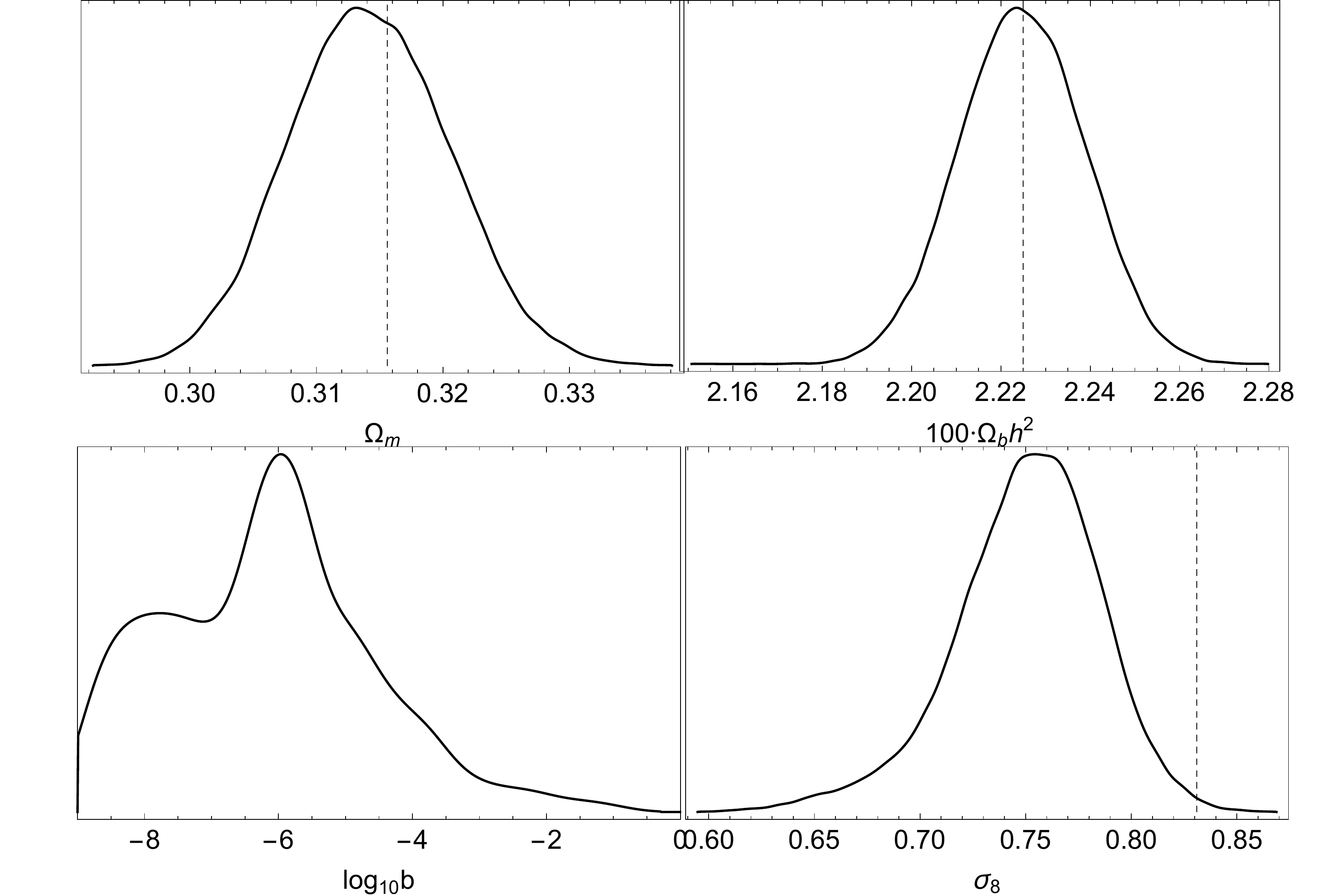}
\caption{The 68.3$\%$, 95.4$\%$ and 99.7$\%$ confidence contours (top) and the one-dimensional marginalized likelihoods (bottom) for various parameter combinations for the HS model. The red point and black dashed lines correspond to the concordance Planck 2015 $\Lambda$CDM parameters given in Table \ref{tab:planck}. The black point indicates the mean value from the MCMC analysis. \label{fig:MCMCHS}}
\end{figure*}

In Figs.~\ref{fig:MCMCLCDM}, \ref{fig:MCMCDES} and \ref{fig:MCMCHS} we show the 68.3$\%$, 95.4$\%$ and 99.7$\%$ confidence contours for the $\Lambda$CDM, the DES and the HS models, respectively, along with the one-dimensional marginalized likelihoods for various parameter combinations. In these plots we also highlight, with either a red point or a black dashed line, the Planck 2015 concordance cosmology. The latter is based on the TT,TE,EE+lowP spectra, a flat $\Lambda$CDM model and the values are shown in Table \ref{tab:planck}. In all cases we find the best-fit $\sigma_{8,0}$ parameter is roughly $\sim 2.5 \sigma$ away from the Planck 2015 best-fit, thus reaffirming the mild tension between low redshift probes and Planck \cite{Nesseris:2017vor}. However, it should be mentioned that there exist several minima in the likelihood with respect to the modified gravity parameters $f_{R,0}$ and $b$ due to the presence of degeneracies in the growth factor, something which has already been studied in standard GR DE models in Ref.~\cite{Nesseris:2011pc}.

Furthermore, we find that a mild tension between Planck and low redshift probes remains even in the case of the $f(R)$ models since in general these cannot predict a decreasing $G_{eff}$ which is required by the growth data, in agreement with Refs.~\cite{Nesseris:2017vor},\cite{Gannouji:2018ncm}. It should be stressed though, that the first year results from the Dark Energy Survey, whose precision is now comparable to that of Planck \cite{Abbott:2017wau}, hints that the tension might be decreasing. Although the central values measured by the Dark Energy Survey for $\sigma_{8,0}$ and $\Omega_{m0}$ are a bit lower compared to those of Planck, it was shown in Ref.~\cite{Abbott:2017wau} that the corresponding Bayes factor are similar; thus, the two datasets are becoming more consistent.

\begin{table*}[!t]
\caption{The best-fit (top row) and mean (bottom row) parameters for the $\Lambda$CDM, the DES and the HS models respectively. Note that $\alpha=(-f_{R,0},b)$. \label{tab:bestfits}}
\begin{centering}
\begin{tabular}{cccccc}
Model & $\Omega_{m0}$ & $100\Omega_b h^2$ & $log_{10}(\alpha)$ & $h$ &$\sigma_{8,0}$ \\\hline
Best-fit values &  &  &  &  &\\\hline
$\Lambda$CDM & $0.313\pm0.006$ & $2.226\pm0.013$ & $-$ & $0.674\pm0.004$ &$0.760\pm0.029$\\\hline
DES & $0.314\pm0.006$ & $2.226\pm0.014$ & $-8.821\pm1.946$ & $0.674\pm0.005$ &$0.753\pm0.043$  \\\hline
HS & $0.315\pm0.006$ & $2.224\pm0.014$ & $-8.186\pm1.510$ & $0.674\pm0.005$ &$0.757\pm0.036$  \\ \hline \hline
Mean values &  &  &  &  &\\\hline
$\Lambda$CDM & $0.314\pm0.006$ & $2.224\pm0.014$ & $-$ & $0.674\pm0.004$ &$0.760\pm0.029$\\\hline
DES & $0.314\pm0.006$ & $2.225\pm0.014$ & $-6.391\pm1.916$ & $0.674\pm0.005$ &$0.738\pm0.043$  \\\hline
HS & $0.314\pm0.006$ & $2.225\pm0.014$ & $-6.176\pm1.567$ & $0.674\pm0.005$ &$0.750\pm0.035$\\\hline
\end{tabular}
\par
\end{centering}
\end{table*}

\begin{table}[!t]
\caption{The $\chi^2$ and AIC parameters for the $\Lambda$CDM, the DES and the HS models respectively. \label{tab:chi2AIC}}
\begin{centering}
\begin{tabular}{cccc}
Model & $\chi^2$ & AIC & $\Delta$AIC \\\hline
$\Lambda$CDM & $1086.62$ &$1094.660$ &$0$ \\\hline
DES & $1086.63$ &$1096.684$ &$2.028$ \\\hline
HS & $1086.61$ &$1096.664$ &$2.008$ \\\hline
\end{tabular}
\par
\end{centering}
\end{table}

In Tables~\ref{tab:bestfits} and \ref{tab:chi2AIC} we show the best-fit, mean values of the model parameter, and also the values for the $\chi^2$ and AIC parameters for the $\Lambda$CDM, the DES and the HS models respectively. As can be seen from Tables \ref{tab:bestfits} and \ref{tab:chi2AIC}, we find that as the difference in the AIC parameters is roughly $\sim 2$, then all three models seem to be statistically consistent with each other. 

\section{Conclusions}
\label{Section:conclusions}

In this paper, we discussed in-depth the effective fluid approach and perturbation theory in the context of $f(R)$ theories. We presented several new results, in particular regarding the effective DE fluid components of the energy momentum tensor, the effective velocity of the fluid $V_{DE}$ given by Eq.~(\ref{eq:efftheta}), the effective pressure and sound speed given by Eqs.~(\ref{eq:effpres}) and (\ref{eq:cs21}). We used these expressions in our modifications of the popular CLASS code, which we call EFCLASS. They provide a much simpler and less error-prone approach in including the effects of modified gravity models.

We then considered specific $f(R)$ models: the well known designer $f(R)$ model (DES), which mimics exactly $\Lambda$CDM at the background level, and the Hu-Sawicki (HS) model which can evade solar system tests. For these models, we calculated the solutions of the DE fluid in the matter dominated era, which we later used as initial conditions for the numerical solution of the system. In this regard, we anticipated the evolution of the numerical solutions by studying the behavior of the DE effective sound speed at both early and late times. As shown, the DE effective sound speed is positive at early times, but then quickly it goes to zero at late times and as a result, the DE perturbations first grow quickly, but then at late times flatten out and reach a plateau. We also found that the numerical solutions of the matter perturbations are in good agreement with the $f\sigma_8$ data and we later on used them in our MCMC analysis. Finally, we also confirmed that for these models the Strong Energy Condition (SEC) is violated, in agreement with the expectation for an accelerating Universe.

With these at hand, we then presented EFCLASS, namely our modifications of the CLASS code, and compared it with other codes in the literature, such as EFTCAMB, CLASS\_EOS\_FR and FRCAMB. The differences between our modifications, discussed in Appendix \ref{Section:class-implementation}, are twofold. First, in contrast to other codes we treat the background of the $f(R)$ models properly by including the correct evolution of the Hubble parameter. In particular, in the case of the HS model we implement very accurate (better than $<10^{-5} \%$) second order analytic approximations for the Hubble parameter $H(z)$. Second, our modifications are overall much simpler and less error-prone than the ones found in other codes, as we use the effective fluid approach variables, namely the effective velocity of the fluid $V_{DE}$ given by Eq.~(\ref{eq:efftheta}) and the anisotropic stress given by Eqs.~(\ref{eq:effpi}). As a result, since we also properly modify the background in the case of $f(R)$ model, we clearly go beyond the simple comparison of Boltzmann codes as was done in Ref.~\cite{Bellini:2017avd}. While for the DES model we find that our results are in good agreement with expectations and other codes, we find a big difference in the case of the HS model, as the other codes currently ignore the necessary modifications to the background.

An important and related issue is also that the viscosity parameter $c_{vis}^2$ actually is not constant as commonly assumed, but rather evolves significantly, as shown in Fig.~\ref{fig:viscocity} where we can see the parameter change by more than 7 orders of magnitude over the range $a\in[10^{-3},1]$. This means that in realistic models, like the Hu-Sawicki $f(R)$ model, $c_{vis}^2$ clearly cannot be considered as a constant parameter, as is the usual assumption when performing forecasts for future surveys, something which in the future should be taken into account.

Finally, we also presented results from our MCMC analysis using the latest cosmological probes including  SnIa, BAO, CMB, $H(z)$ and growth $f\sigma_8$ data. We presented a complete analysis and a Bayesian comparison of the $\Lambda$CDM, DES and HS models. The confidence contours and one-dimensional marginalized likelihoods from the MCMC analysis were shown in Figs.~\ref{fig:MCMCLCDM}, \ref{fig:MCMCDES} and \ref{fig:MCMCHS}, while in Tables~\ref{tab:bestfits} and \ref{tab:chi2AIC} we showed the best-fit, mean values of the model parameters, but also the values for the $\chi^2$ and AIC parameters for the $\Lambda$CDM, the DES and the HS models respectively. We found that as the difference in the AIC parameters is roughly $\sim 2$, then all three models can be assumed to be statistically consistent with each other.

To summarize, we showed that by using our new expressions for the DE effective fluid description of the $f(R)$ models as described earlier and the simple modifications to the CLASS code in conjunction to the very accurate analytic approximations for the background evolution, we can obtain competitive results in a much simpler and less error-prone approach. In particular, the correct treatment of the background evolution is very important, as in the near future we will have access to cosmological data that constrain the background to less than 1 percent, thus our theoretical predictions must also be at least as accurate.


\textbf{Numerical Analysis Files}: The numerical codes (Fortran, C, Mathematica and Python) used by the authors in the analysis of the paper and our modifications to the CLASS code, which we call EFCLASS, can be found \href{https://members.ift.uam-csic.es/savvas.nesseris/efclass.html}{here} and \href{https://github.com/wilmarcardonac/EFCLASS}{here}.

\section*{Acknowledgements}
The authors would like to thank G.~Ballesteros, J.~Garc\'{\i}a-Bellido, M.~Kunz, J.~Lesgourgues, A.~Maroto, F.~Montanari, L.~Pogosian, D.~Sapone, I.~Sawicki and A.~Silvestri for many fruitful discussions. The authors acknowledge support from the Research Project FPA2015-68048-03-3P [MINECO-FEDER], the Centro de Excelencia Severo Ochoa Program SEV-2016-0597 and use of the Hydra cluster at the IFT. S.N. also acknowledges support from the Ram\'{o}n y Cajal program through Grant No. RYC-2014-15843.

\appendix
\section{Useful formulae and ISW effect}
\label{Section:useful-formulae}

In this section we present some useful formulas related to the effective fluid approach and the ISW effect. Using the definitions of the effective pressure perturbation, the anisotropic stress  and the effective sound speed one can easily obtain the following expressions:
\ba
\delta P_{DE}&=&\frac13 T, \\
\Sigma^i_j&=& T^i_j-\frac13\delta^i_j T, \\
(\bar{\rho}+\bar{P})\sigma &=& -(\hat{k}_i \hat{k}_j-\frac13 \delta_{ij}) \Sigma^{ij},\\
\pi_{DE} &=& \frac32(1+w)\sigma, \\
c_{s,eff}^2 \delta \rho_{DE}&=& \delta P_{DE}-\frac23 \bar{\rho}_{DE} \pi_{DE},
\ea
which lead to
\be
\bar{\rho}_{DE} \pi_{DE}=-\frac32 \left(\hat{k}_i \hat{k}_j T^{ij}-\frac{T}{3}\right)
\ee
and
\be
c_{s,eff}^2 \delta \rho_{DE}=\hat{k}_i \hat{k}_j T^{ij}
\ee
where $T=T^i_i$, $\hat{k}_i$ is a unit vector in Fourier space and in the above expressions we have only kept the 1st order parts.

In what follows we present the theoretical expressions used to calculate the low multipoles for Fig.~\ref{fig:cmbclsISWtheory}. In this regard, we mostly follow Ref.~\cite{Song:2006ej}. The contribution of the ISW effect on the angular CMB power spectrum is given by \cite{Song:2006ej}:
\bea
C_\ell^{\textrm{ISW}}=4\pi \int \frac{dk}{k} I_\ell^{\textrm{ISW}}(k)^2 \frac{9}{25} \frac{k^3 P_{\zeta}}{2\pi^2},\label{eq:clsISWtheory}
\eea
where we have used the fact the power spectrum $P_{\zeta}$ is given in terms of the primordial power spectrum  times a transfer function
\be
\frac{k^3 P_{\zeta}}{2\pi^2}=A_s \left(\frac{k}{k_0}\right)^{n_s-1} T(k)^2,
\ee
where $A_s$ is the primordial amplitude, $k_0$ is the pivot scale and $T(k)$ is the usual matter-radiation transfer function (see Eq. (7.71) in Ref.~\cite{Dodelson:2003ft}). Furthermore, the kernel $I_\ell^{\textrm{ISW}}(k)$ is given by

\be
I_\ell^{\textrm{ISW}}(k)=2\int dz \frac{dG}{dz} j_\ell (k\;r(z)),
\ee
where $j_n(x)$ is the spherical bessel function, $r(z)=\int_0^z dz/H(z)$ is the comoving distance and the function $G(z,k)$ is the scale dependent potential growth rate
\be
G(a,k)=\frac{\Phi(a,k)+\Psi(a,k)}{\Phi(a_{ini},k)+\Psi(a_{ini},k)}.
\ee
Also, the contribution to the spectrum due to the usual Sachs-Wolfe (SW) effect is given by:
\be
C_\ell^{\textrm{SW}}=\frac{2 \pi}{25}A_s \frac{\Gamma\left(\frac32\right)\Gamma\left(1-\frac{n_s-1}{2}\right)\Gamma\left(\ell+\frac{n_s-1}{2}\right)}{\Gamma\left(\frac32-\frac{n_s-1}{2}\right)\Gamma\left(\ell+2-\frac{n_s-1}{2}\right)},\label{eq:clsSWtheory}
\ee
where $\Gamma(x)$ is the usual Gamma function. The previous expression for $n_s=1$ simplifies to the well-known result for the SW plateau
\be
\frac{\ell(\ell+1)}{2\pi} C_\ell^{\textrm{SW}}=\frac{A_s}{25}.
\ee

Finally, the total contribution from the SW and ISW effects will be given by the sum of Eqs.~\eqref{eq:clsISWtheory} and \eqref{eq:clsSWtheory}, that is,
\be
C_\ell^{\textrm{total}}=C_\ell^{\textrm{SW}}+C_\ell^{\textrm{ISW}}.\label{eq:clsalltheory}
\ee
In our analysis we used $A_s=2.3\times 10^{-9}$, $n_s=1$, $k_0=0.05 h/\textrm{Mpc}$, $\Omega_{m0}=0.3$ and $T_{\textrm{CMB}}=2.726K$. Note that to convert the result of Eq.~\eqref{eq:clsalltheory} to $\mu K^2$, as is the standard in the CMB community, one needs to multiply the $C_\ell$ with $T_{\textrm{CMB}}^2\cdot 10^{12}$. 

\section{CLASS implementation}
\label{Section:class-implementation}

In this section we present our implementation of the effective fluid approach in the CLASS code \cite{Blas:2011rf}, which we call EFCLASS. As shown in the previous sections, even with these minimal changes our approach gives results in agreement with other codes, such as EFTCAMB, MGCAMB, FRCAMB and CLASS\_EOS\_FR.

The only changes we made in the code are in the following two places:
\begin{enumerate}
  \item In the \emph{background.c} file we included the correct expansion history for the $f(R)$ models. For the HS model this is given by Eq.~\eqref{eq:HubHSapp}.
  \item In the \emph{perturbations.c} file we included the proper perturbations for the effective DE fluid given by Eqs.~(\ref{eq:phiprimeeq1}) and (\ref{eq:anisoeq}).
\end{enumerate}

We found that the most straightforward and least error-prone way to make these changes is to modify the $\Lambda$CDM model equations in the aforementioned parts of the code. First, we can just increment the background equations of $\Lambda$CDM with the one of the HS model (for the DES model, no change is needed). Second, since $\Lambda$CDM has no perturbations we can just add the appropriate new terms given by Eqs.~(\ref{eq:phiprimeeq1}) and (\ref{eq:anisoeq}).

In more detail, first we consider the background evolution, where we consider two cases: that of the DES model, where the background is fixed to that of the $\Lambda$CDM model, and that of the HS model where the Friedman equation is modified. For the DES model we obviously do not make any change as the Hubble parameter for the $\Lambda$CDM is already included in the CLASS code. For the HS model we introduce the extremely accurate approximations for the Hubble parameter given by Eq.~(\ref{eq:HubHSapp}). In Ref.~\cite{Basilakos:2013nfa} is shown that this expression works to a level of accuracy better than $\sim10^{-5}\%$ for $b\in[0,0.1]$. Finally, we also had to include an expression for the equation of state parameter $w_{DE}$ and effective density $\rho_{DE}$. Both were calculated to second order in $b$ from Eqs.~(\ref{eq:wde}) and  (\ref{eq:effden}) by using Eq.~(\ref{eq:HubHSapp}).

Regarding the perturbations, we treat both models equally. In this case we found that the best place to implement the modifications were in the \emph{perturb\_einstein} routine of CLASS, which solves the Einstein equations in the conformal Newtonian gauge given by Eqs.~(\ref{eq:phiprimeeq1}) and (\ref{eq:anisoeq}). Then, it is simple to just add in the right-hand-side of the aforementioned equations our expressions for the effective fluid DE velocity and anisotropic stress given by Eqs.~(\ref{eq:efftheta}) and (\ref{eq:effpi}).

Our analytic approach has several advantages. First, given that most viable $f(R)$ models can be written as small perturbations around $\Lambda$CDM model, such as the HS model, it is always possible to derive extremely accurate expressions for the background, as was shown in Ref.~\cite{Basilakos:2013nfa}. Second, regarding the perturbations our improved sub-horizon approximation gives much more accurate results compared to codes that are based on the default sub-horizon approximation. Also, the accuracy is comparable to codes that treat the perturbations exactly by numerically solving the relevant equations. However, our approach has a much smaller overhead in terms of new lines of code and as a result is more straight-forward and less error-prone. 

\bibliography{effective_fluids}

\end{document}